\documentclass[11pt,a4paper]{article}
\usepackage[utf8]{inputenc}
\usepackage[english]{babel}
\usepackage{amsmath}
\usepackage{amsfonts}
\usepackage{amssymb}
\usepackage{makeidx}
\usepackage{graphicx}
\usepackage{lmodern}
\usepackage{kpfonts}
\usepackage{multirow}
\usepackage{supertabular}
\usepackage{dcolumn}
\usepackage{pdflscape}
\usepackage{pgfplots}
\usepackage{accents}
\usepackage{setspace}
\usepackage{booktabs}
\usepackage{longtable}
\usepackage[round]{natbib}
\usepackage[left=2cm,right=2cm,top=2cm,bottom=2cm]{geometry}

\begin{document}

{\center {\Large Swimming in Dark Water: When Cartels Mimic Competition}{\large
\vspace{0.1cm}}\smallskip\\

{\large David Imhof*, Thierry Madies**, and Martin Huber**}\smallskip\\
{\small {* Swiss Competition Commission, University of Fribourg, Dept.\ of Economics and Unidistance (Switzerland).}}\\[0pt]
{\small {** University of Fribourg, Dept.\ of Economics}}\\[0pt]}\smallskip

\vspace{2cm} \noindent \textbf{Abstract:} {\small \textit{This paper analyzes the internal organization and economic effects of a bid-rigging cartel in the road construction sector of the Swiss canton of Ticino, active from 1999 to 2005. Using exceptionally rich documentary evidence, we reconstruct how cartel members coordinated bids and allocated contracts under a formal agreement known as the “convention”.
We show that, despite the absence of side payments, the cartel implemented a cost-based allocation mechanism that closely approximated the first-best collusive outcome. Regression and machine-learning analyses indicate that observable cost proxies systematically predict both winning bids and bid rankings. The evidence further suggests that cartel members strategically mimicked competitive bidding behavior, allowing them to evade standard econometric detection methods.
Using double machine learning, we estimate average overcharges of at least 45\%, and potentially substantially higher, highlighting the significant financial harm caused by this sophisticated form of collusion.}}
\vspace{0cm}\smallskip\\

\vspace{2pt}

\vspace{2pt}

{\footnotesize 

\noindent Addresses for correspondence:  David Imhof, Hallwylstrasse 4, 3003 Bern, Switzerland; david.imhof5@gmail.com. Thierry Madies, University of Fribourg, Bd.\ de P\'{e}rolles 90, 1700 Fribourg, Switzerland; thierry.madies@unifr.ch. Martin Huber, University of Fribourg, Bd.\ de P\'{e}rolles 90, 1700 Fribourg, Switzerland; martin.huber@unifr.ch.}\vspace{2pt}

{\footnotesize \noindent \textbf{Disclaimer:} All views contained in this paper are solely those of the authors and cannot be attributed to the Swiss Competition Commission, its Secretariat, the University of Fribourg or Unidistance (Switzerland).}

\thispagestyle{empty}\pagebreak  {\small
\renewcommand{\thefootnote}{\arabic{footnote}}
\setcounter{footnote}{0} \pagebreak \setcounter{footnote}{0}
\pagebreak \setcounter{page}{1} }

\newpage

\section{Introduction}
Public procurement represents a substantial share of economic activity, accounting for about 13\% of GDP in OECD countries \citep{OECD2016}. Bid-rigging cartels in this domain inflict significant financial harm by artificially inflating prices, with overcharges commonly estimated between 15\% and 30\% \citep{Bolotova2008,Froeb1993,connor2014}. The economic relevance of this phenomenon is heightened by the fact that the probability of detection and conviction remains low, around 10–20\% per year according to several studies \citep{BryantEckard1991,Combe2008,HarringtonWei2017}. Against this background, understanding how cartels are able to coordinate successfully while remaining hidden is of both academic and policy relevance.

In this study, we examine whether and how a long-running cartel in the Swiss canton of Ticino operated according to an internally consistent allocative mechanism maximizing the cartel rent while successfully evading detection by the procurement agency and the competition authority. In other words, we ask if a “weak” cartel, defined as a cartel operating without side payments, can replicate the allocative efficiency of a “strong” cartel by assigning contracts to the lowest-cost bidder and aligning the distribution of bids with underlying cost structures, thereby mimicking competitive outcomes. To the best of our knowledge, this question has not been addressed empirically in the literature, despite its importance for both the theoretical understanding of collusion and the design of effective antitrust policies.

The case studied in this paper is exceptional because the cartel was formalized in a written agreement of thirteen articles, called the convention, organizing the cartel’s operation in a remarkably detailed manner. All public and private contracts had to be announced and discussed during mandatory weekly meetings. The convention codified the allocation rules and distributed contracts based on a set of cost-related and fairness criteria: backlog of ongoing contracts, distance to the contract site, firm specialization, and collegial discussions. Voting procedures resolved conflicts, while firms were obliged to prepare cover bids to secure the cartel’s designated allocation. No side payments were used; instead, the written agreement and the threat of exclusion from access to inputs were sufficient to ensure compliance. The unusually rich documentation of the case allows for an in-depth empirical analysis of the cartel's functioning.

Against this background, our empirical analysis pursues three objectives. First, we investigate whether the Ticino cartel effectively functioned as a weak cartel (without side payments) capable of replicating the allocation efficiency of a strong cartel. According to theory \citep{McAfee1992,Athey2001,Marshall2007,Pesendorfer2000}, weak cartels are generally less stable and less efficient but may attain the first-best collusive outcome if contracts are sufficiently numerous and a ranking mechanism ensures truthful revelation of costs. Empirical evidence further suggests that cartels often rely on sophisticated allocation and coordination rules to sustain collusion \citep{Genesove2001,Asker2010,Ishii2009}. The Ticino convention appears to provide such a mechanism. We test this hypothesis by analyzing whether cost proxies, such as distance and backlog, systematically predict the winning bids and the ranking of all bids within tenders.

Second, we assess how closely the cartel's behavior mimicked competitive conduct. Using econometric tests developed by \citeauthor{Bajari2003} \citeyearpar{Bajari2003}, we examine whether bids are conditionally independent and exchangeable given cost proxies. In competitive settings, these conditions should hold. Our analysis, however, reveals that the Ticino cartel managed to align bids with cost structures so effectively that standard econometric tests generated a high rate of false negatives. In other words, the cartel was able to pass as competitive while being fully collusive.

Third, we estimate the damage caused by the cartel. We apply both traditional econometric models and modern machine learning methods, namely lasso regression \citep{Tibshirani1996}, ensemble approaches \citep{Breiman2001,Zhou2012}, and double machine learning \citep{Chetal2018,DoubleML}, in order to quantify price overcharges.
Our findings indicate that the Ticino cartel increased procurement prices by about 45\% or more, which is substantially higher than the average overcharge estimates reported in the existing literature \citep{Bolotova2008, connor2014}.
These results underline the considerable harm inflicted on taxpayers and highlight the need for more effective detection methods. They also illustrate how well the cartel managed to maximize the first-best collusive outcome despite its hypothetical weakness.

Our paper contributes to the literature along three main dimensions. First, this empirical study is able to show that a weak cartel, i.e. a cartel operating without monetary transfers, can nevertheless achieve the first-best collusive outcome. Previous studies have either focused on strong cartels with side payments \citep{Asker2010} or documented weak cartels that failed to extract maximum rents \citep{Ishii2009}.

Second, it raises methodological implications for cartel detection. Building on \citeauthor{Bajari2003} \citeyearpar{Bajari2003} and related works \citep{Porter1993,Porter1999,Ishii2009,Aryal2013,Lundberg2025}, we show that econometric tests based on cost proxies may fail to detect collusion when cartels internalize cost logic in their bidding behavior. The Ticino cartel deliberately aligned bids with observable cost structures, thereby mimicking competitive conduct and escaping detection by cost-based methods. This result warns against over-reliance on a single type of detection method and suggests the need for complementary approaches, \textit{e.g.} including machine learning methods capable of uncovering subtle collusive patterns in the distribution of bids.

Third, our paper advances research on cartel damage estimation. Our damage estimates exceed those reported in many earlier studies \citep{Bolotova2008,connor2014}. By combining flexible machine learning methods with detailed data, we provide robust evidence of very large overcharges in this market. This not only contributes to the literature on cartel damage but also offers practical insights for competition authorities seeking to quantify harm in procurement cases.

To sum up, the study of Ticino cartel sheds new light on how weak cartels can replicate the efficiency of strong cartels, how collusion can remain undetectable despite sophisticated econometric tests, and how damaging such collusion can be for public finances. More broadly, our findings underscore the importance of studying the internal organization of cartels using detailed documentation, as this may reveal coordination mechanisms that allow cartels both to maximize rents and to elude detection.

The remainder of the paper is organized as follows. Section 2 introduces the Ticino cartel and its institutional context. Section 3 presents the key variables in the data and outlines the models and the estimation strategy.
Section 4 presents the empirical results for the logit and ordered-logit estimations, as well as the econometric tests based on cost-related variables.
Section 5 provides estimates of the average effect of the Ticino cartel on prices to assess cartel damage. Section 6 offers policy recommendations and section 7 concludes.

\section{The Ticino Cartel}
This section describes the institutional environment and internal organization of the Ticino bid-rigging cartel. All firms active in the road construction sector in the Canton of Ticino colluded on virtually every contract between January 1999 and March 2005.\footnote{One firm did not participate in the cartel. However, because it does not appear in our dataset, we assume it was a small company not engaged in public road construction procurement.} The agreement was formalized in a written document of thirteen articles, known as the convention, which governed both the structure and the day-to-day operation of the cartel.\footnote{See Appendix \ref{AppendixA} for a translation in English of the convention of the Ticino bid-rigging cartel in decision Strassenbeläge Tessin (LPC 2008/1, pp. 78–80).}
Weekly meetings constituted the core coordination device. During these meetings, cartel members examined all public tenders and all private contracts exceeding CHF 20,000, and allocated them among participants according to predefined rules. Once a designated winner was selected, firms coordinated bid prices to ensure that the agreed allocation was implemented in the procurement process. In what follows, we describe the market structure, the allocation mechanism, the criteria specified in the convention, and the implications of price coordination for competition and pricing.

\subsection{A market prone to collusion}
The Swiss Competition Commission (COMCO) defined two relevant markets in its investigation: the downstream market for road construction and pavement in the Canton of Ticino, and an upstream market for asphalt pavement materials. Seventeen firms were active in road construction, while two firms (firm A and firm B) operated asphalt mixing plants supplying the downstream market. Asphalt is a essential input for road construction: it must be heated at a mixing plant, transported quickly, and applied before cooling. Market experts estimate a workable time window of roughly one to one and a half hours, implying an operational radius of approximately 50–80 km around a mixing plant. These technological constraints make asphalt production a strategic input and confer substantial market power on upstream suppliers.

Among the construction firms, four (firms 3, 4, 5, and 6 in Table \ref{Tabledesc3} in Appendix \ref{AppendixA}) were vertically integrated and owned mixing plants. Because asphalt production requires large fixed investments, several firms pooled resources to jointly own production facilities. The largest plant (firm A) was co-owned by ten firms and accounted for roughly 50–60\% of asphalt supply in Ticino. The second-largest plant (firm B), jointly owned by firms 4 and 16, held a market share of about 10–20\%. Firm A also maintained a financial participation in firm B. In total, twelve road construction firms jointly controlled between 60\% and 80\% of the upstream asphalt market.

The cartel convention explicitly prohibited the sale of asphalt and other essential inputs to firms outside the agreement.\footnote{See Art. 6 Obligation.} This rule effectively foreclosed the downstream market and created strong entry barriers. Any potential entrant without access to asphalt would have been forced to build a mixing plant, an investment sufficiently costly to deter entry. Compliance was enforced through severe disciplinary measures: firms deviating from the agreement risked being cut off from asphalt supplies and thus excluded from the road construction market altogether.

Beyond entry barriers, several structural features further facilitated collusion. The OECD identifies a number of conditions that make markets particularly prone to coordination \citep[see][]{OECD2016}. Table \ref{structuralscreench1} summarizes these factors for Ticino. Although market concentration was moderate (17 active firms), contract values were highly asymmetric, ranging from CHF 8 to 26 million (Table \ref{Tabledesc3}). Such asymmetries can, in principle, weaken collusion by increasing incentives to deviate \citep[see][]{Brock1985, Compte2003}.

However, other forces strongly supported cartel stability. Firms interacted repeatedly, and the bidding process was highly transparent: bid summaries were observable to all participants. Repeated interaction and transparency are well-known facilitators of collusion \citep[see][]{Green1984, Bernheim1990, Hendricks2015}. Innovation in road construction was essentially absent, limiting competitive disruption \citep[see][]{Ivaldi2003}. Cost structures were similar across firms, easing coordination \citep[see][]{Rey2006}, whereas cross-shareholdings in the upstream asphalt market reinforced incentives to collude \citep[see][]{Gilo2006}. On the demand side, procurement authorities were the dominant buyers, with a relatively inelastic demand and predictable timing of tenders. Together, these features created an environment highly conducive to stable, long-run collusion.

To sum up, the Ticino market combined technological, institutional, and organizational elements that made cartelization both feasible and sustainable.

\begin{table} [!htp]
\begin{center}
\begin{tabular}{lc}\hline\hline
Market Characteristics&Collusive Assessment for the Ticino Case\\
Market Concentration&(-)\\
Entry barriers&(+)\\
High frequency of interaction between competitors&(+)\\
Market Transparency&(+)\\
Mature Industry with little innovation&(+)\\
Similar cost Structure&(+)\\
Asymmetric Distribution of Production Capacities&(-)\\
Homogenous Product&(+)\\
Cross-shareholdings&(+)\\
Demand Fluctuation&(+)\\
Elasticity of the Demand&(+)\\
\hline
\end{tabular}
\end{center}
\caption{Structural screens } \label{structuralscreench1}
\end{table}

\subsection{The contract allocation mechanism}
The cartel convention consisted of three pages and thirteen articles.\footnote{See the convention in Appendix \ref{AppendixA}.} It mandated weekly meetings that all firms were required to attend. Unjustified absences were sanctioned and could result in the loss of future contracts. At each meeting, participants had to announce all new public procurement tenders and all private construction contracts above CHF 20,000.\footnote{See convention, art. 5 Scope.} Allocation and pricing strategies were then discussed collectively.\footnote{See convention, arts. 4 Organization and 7 Allocation of Contracts.}

According to art. 7 of the convention, contract allocation depended on the following criteria:
\begin{description}
\item[$\ast$ (a)] Amount of contracts previously won
\item[$\ast$ (b)] Location of the contract
\item[$\ast$ (c)] Specialization of each firm
\item[$\ast$ (d)] Private bidding
\item[$\ast$ (e)] Collegial discussion among cartel members
\end{description}

The first criterion prioritized firms with the smallest backlog of ongoing work. The second favored firms located closest to the project site, particularly for contracts below CHF 500,000. Both criteria align closely with cost proxies commonly used in structural bidding models \citep[see][]{Porter1993, Porter1999, Pesendorfer2000, Bajari2003, Ishii2009, Aryal2013, Chotibhongs2012a, Chotibhongs2012b, Lundberg2025}. From an economic perspective, firms with lower workload and shorter transport distances are likely to have lower marginal costs, making these rules consistent with an efficiency-oriented allocation of contracts within the cartel.

The third criterion concerned specialization. While intuitively reasonable, detailed data on firm specialization are unavailable, and contracts in road construction are relatively homogeneous. As a result, it is difficult to operationalize this criterion empirically.

The fourth criterion gave priority to the firm that first prepared a quotation for a private client.\footnote{The convention also considered private contracts above 20'000 CHF for allocation among cartel participants.} Preparing bids for small private contracts involved sunk costs because clients typically did not provide bidding documentation. Granting priority to the first firm preparing an offer thus incentivized cartel members to disclose private contracts during meetings. This rule helped the cartel maintain accurate information about each firm's workload, as private contracts were more difficult to monitor than public tenders.

Finally, collegial discussion served a fairness objective rather than cost efficiency, functioning similarly to a simple bid-rotation scheme. Final decisions for allocating contracts among cartel members were made by majority vote, and in case of disagreement a secret ballot was organized, excluding firms involved in litigation.

\subsection{Price coordination}

After allocating contracts, cartel members coordinated bid prices. The convention required firms to calculate bids prior to meetings and then adjust them collectively.\footnote{See convention, art. 6 Scope.} Firms whose initial bids were below the agreed cartel price had to inflate them artificially to protect the designated winner.
These so-called cover bids had to be formally prepared and fully justifiable using official tender documentation.\footnote{See convention, art. 4 Organization.} At the same time, they had to remain sufficiently high not to threaten the ranking. This ensured that procurement agencies would disregard them in the award decision, thereby implementing the cartel's allocation.

The cartel operated under virtually no competitive discipline. Deviations were deterred by credible sanctions, most notably exclusion from access to asphalt and other key inputs. Moreover, COMCO defined the relevant geographic market as confined to Ticino, isolated by political and natural barriers. Access to mixing plants was controlled by cartel members, further limiting outside entry. The cartel therefore faced little external pressure when setting prices.

This absence of competition translated into substantial price increases. COMCO compared engineer estimates with bids before and after the cartel collapsed. After March 2005, lowest bids were roughly 30 percentage points below estimates, whereas during the cartel period bids closely tracked or exceeded them.\footnote{See decision Strassenbeläge Tessin, LPC 2008-1, p. 103, Figure 1}
This shows that engineers had internalized cartel prices when forming benchmarks, consistent with the endogenous adjustment mechanism described by \citet{Harrington2006}.

\section{Key Variables and Estimation Strategies}
\subsection{The procurement process and data features}
Procurement agencies in Ticino announced a fixed deadline for the submission of bids for each contract and provided all technical and administrative documents required for participation. Interested firms prepared and submitted their bids before the deadline. After the submission period closes, procurement agencies opened and recorded all bids in official bid summaries, which reported bid values, bidder identities, contract locations, and contract types.

After recording, agencies evaluated bids according to several criteria, including firms’ references, project organization, timing, and social or environmental considerations. Nevertheless, price remained the dominant criterion. In practice, contracts are typically awarded to the lowest bid. During the cartel period, all winning bids were indeed the lowest bids in each tender, implying that procurement in Ticino operated as a first-price sealed-bid auction.

As shown in Table \ref{Tabledesc1}, our dataset consists of 183 collusive tenders from January 1999 to March 2005 and 40 competitive tenders from April 2005 to April 2006. These correspond to 1,060 collusive bids and 229 competitive bids. Consortia account for 17\% of bids during the cartel period and 23\% in the post-cartel period.\footnote{Firms are allowed to bid jointly for a specific contract. When bidding together, they form a consortium, sharing risks and benefits according to an agreement that defines each participant's share.}

{\renewcommand{\arraystretch}{1.1}
\begin{table} [!htp]
\begin{center}
\begin{tabular}{rrr}
\hline \hline
Period & Cartel & Post-cartel\\
\hline
Tenders & 183 & 40 \\
Firms & 17 & 16 \\
Bids & 1060 & 229 \\
Bids in consortium & 180 & 42 \\
Individual bids & 880 & 187 \\
Sales in mio. CHF & 248.2 & 34.0 \\
\hline
\end{tabular}
\end{center}
\caption{Descriptive statistics} \label{Tabledesc1}
\end{table}}

Table \ref{contractsch1} reports annual contract volumes tendered in Ticino. Total procurement varies between CHF 19 and 52 million. This fluctuation reflects the institutional structure of procurement: the canton organized large maintenance tenders every two years, and these contracts typically lasted two years.

Table \ref{Tabledesc3} in Appendix \ref{AppendixA} reports firm-level statistics during the cartel period and illustrates substantial heterogeneity in participation, bidding activity, and success rates. Some firms submitted bids frequently but won few contracts, while others participated less often but secured a large share of projects. For instance, firm 12 won contracts totaling CHF 21.6 million, whereas firms 6 and 16 won roughly CHF 10 million each. Firm 4 submitted 119 bids but won only 15 contracts, while firm 16 submitted 29 bids and won 11 tenders. These patterns highlight the dispersion in bidding behavior of each firm and outcomes.

{\renewcommand{\arraystretch}{1.1}
\begin{table} [!htp]
\begin{center}
\begin{tabular}{ccc}\hline\hline
Year&Contracts&Amount in Mio.\\
1999 & 28 & 51.9 \\
2000 & 27 & 31.5 \\
2001 & 24 & 46.8 \\
2002 & 30 & 38.7 \\
2003 & 21 & 39.0 \\
2004 & 45 & 35.3 \\
2005 & 34 & 20.9 \\
2006 & 14 & 19.1 \\
Total& 223 & 283.2 \\
\hline
\end{tabular}
\end{center}
\caption{Number and value of annual tenders in Ticino (CHF)} \label{contractsch1}
\end{table}}

Table \ref{Tabledesc2} displays the distribution of bids per tender. The number of bids does not necessarily coincide with the number of bidders because several firms may submit joint bids through consortia, especially for larger contracts. Across the full sample, tenders typically attract between three and eight bids.

{\renewcommand{\arraystretch}{1.1}
\begin{table} [!htp]
\begin{center}
\begin{tabular}{rrrrrrrrrrrrr}
\hline \hline
Bids by tender&2 & 3 & 4 & 5 & 6 & 7 & 8 & 9 & 10 & 11 & 12 & 13 \\ 
  \hline
Nbr. tenders&13 & 32 & 37 & 25 & 33 & 31 & 22 & 13 & 7 & 6 & 2 & 2 \\ 
   \hline
\end{tabular}
\end{center}
\caption{Number of bids per tender} \label{Tabledesc2}
\end{table}}

\newpage

\subsection{Dependent variables and models}
We now describe the estimation strategies used to assess whether the Ticino cartel behaved as a “weak” or “strong” cartel. A cartel functioning through contract allocation can nevertheless be considered strong if it uses an internal ranking mechanism capable of selecting the lowest-cost bidder \citep[see][]{Pesendorfer2000, Aoyagi2007, Hendricks2015}. Such a mechanism allows the cartel to approximate the first-best collusive outcome even in the absence of explicit monetary transfers. As discussed in the previous section, the Ticino convention may have played this role because several allocation criteria are closely related to the costs of each bidder.

To test whether the cartel selected the lowest-cost bidder, we estimate binary response models in which the dependent variable equals one if a bid is the lowest in a tender and zero otherwise. We denote this variable by \textit{LWTBID}. Using a logistic specification, the conditional probability of submitting the lowest bid is

\begin{equation}\label{eqbinary}
\Pr \left( \text{\textit{LWTBID}} = 1 \mid X \right) = \Lambda(\beta_{0} + \beta_{1}x_{1} + \beta_{2}x_{2} + \dots + \beta_{k}x_{k}) = \Lambda\left(\beta_{0} + X \beta\right),
\end{equation}

where $\Lambda(\cdot)=\frac{\exp(\cdot)}{1+\exp(\cdot)}$ is the logistic link function, $X$ is a vector of explanatory variables, and $\beta$ is the associated coefficient vector. The regressors include proxies for costs, variables capturing contract allocation rules, and firm-level controls such as indicators for major bidders, vertical integration, and ownership links to mixing plants A and B (Table \ref{Tabledesc3} in Appendix \ref{AppendixA}).

We estimate several nested specifications. Model 1 includes only cost proxies. Model 2 includes only variables associated with rotational allocation of contracts. Model 3 combines cost and allocation variables. Model 4 adds indicators for vertical integration and asphalt plant ownership. Model 5 includes firm fixed indicators for the twelve most active bidders (those submitting at least 20 bids). Model 6 includes the full set of cost, allocation, and control variables.

Next, we analyze whether the cartel also coordinated the structure of cover bids in a cost-consistent manner. According to the convention, cover bids should remain justifiable in light of firms' actual cost positions. We therefore construct the dependent variable \textit{RANK}, which denotes a firm's position in the bid distribution of a tender. In a tender with eight bids, the rank ranges from 1 (lowest bid) to 8 (highest bid). For tenders with more than eight bids, we truncate the rank at 8, as such cases are rare (Table \ref{Tabledesc2}).

Because bid positions are ordinal, we estimate ordered logit models rather than multinomial models, following models 3 through 6. Higher ranks are interpreted as reflecting higher underlying costs due, for example, to distance or capacity constraints.

\subsection{Proxies for the costs}
Distance and capacity are standard determinants of bidding behavior in procurement auctions. Greater distance from the project site raises transport and logistical costs, while limited available capacity increases opportunity costs \citep[see][]{Porter1993, Porter1999, Pesendorfer2000, Bajari2003, Aryal2013, Ishii2009, Chotibhongs2012a, Chotibhongs2012b, Lundberg2025}.

We construct \textit{LNDIST} as $LNDIST_{it}=\ln(km_{it}+1)$, where $km_{it}$ is the Google Maps distance between firm $i$'s headquarters and the contract location in tender $t$.\footnote{The km equals zero when the contract location is in the same municipality as the firm's location.} For contracts involving multiple locations, we compute the average distance. For biennial regional maintenance contracts, we identify three key locations and use the average distance to those points.
Capacity is measured by \textit{LNCAP}, defined as the logarithm of the value of contracts won by a firm over the two years preceding the submission of bid $t$. We calculate the capacity over two years, precisely because the Canton of Ticino tendered biennial maintenance contracts of important value, as illustrated by the fluctuation of annual contracts in Table \ref{contractsch1}. Firms with higher backlogs face higher marginal costs and should therefore bid less aggressively.

We also include variables capturing strategic interactions. The first is the minimum rival distance $LNDIST\_SEC_{it}$. For the closest bidder, this equals the second-smallest distance; for all others, it equals the smallest distance among competitors. If rivals are located far away, firm $i$ may bid less aggressively.
The second interaction variable is $LNCAP\_SEC_{it}$, defined as the minimum backlog among rivals. If competitors are heavily loaded with existing contracts, firm $i$ can expect weaker competition and raise its bid accordingly.

\subsection{Contract allocation mechanism}
To capture allocation based on bid rotation, we follow \citet{Ishii2009}. The first variable is \textit{LASTWIN}, the number of days since firm $i$ last won a contract. In a rotation scheme, firms that have waited longer should be favored. We therefore expect \textit{LASTWIN} to increase the probability of submitting the lowest bid, while decreasing the bid rank in the ordered logit model.

Second, we consider revenue equalization. The variable \textit{EQUAL} measures the cumulative value of contracts won by each firm since the beginning of the cartel period. If the cartel equalizes revenues, firms with lower cumulative revenue should be more likely to win future contracts. Thus, \textit{EQUAL} should be negatively associated with winning and positively associated with rank under rotation.

Third, we construct a participation score, \textit{SCORE}, defined as the number of cover bids submitted by firm $i$ since its last win. This captures the idea that firms supporting others are rewarded later. We expect \textit{SCORE} to increase the probability of winning and lower the bid rank.

\subsection{Consortia}
A consortium is a group of firms jointly submitting a bid for a specific contract. In Ticino, some consortia are recurring, while others are one-off. We assign unique identifiers to recurring consortia and a value of zero to sporadic ones.

Cost measures for consortia follow Article 7(a) of the convention. When a consortium competes against individual bidders, the convention assigns the minimum cost among its members, reflecting the idea that consortia exploit comparative advantages. We follow this approach, as it is economically consistent and captures potential efficiency gains.

In logit and ordered logit models, each consortium member remains an observation and receives the same cost value but retains firm-specific allocation variables. Members of the same consortium have the same rank. Robustness checks compare results with and without consortia.
For econometric detection tests based on reduced-form bidding functions, redundant consortium observations are suppressed. Six regular consortia receive identifiers 21 to 26 so that each panel unit appears only once per tender. In these tests, only cost variables are used.

Table \ref{desvarch1} summarizes all variables, and Table \ref{Tabledesc5} in Appendix \ref{AppendixA} reports descriptive statistics for individual bids.

\begin{table} [!htp]
\begin{center}
\begin{tabular}{lp{8cm}}\hline\hline
Variables&Description\\
LWTBID & The variable \textit{LWTBID} takes the value 1 if the bid submitted in tender $t$ is the lowest. \\
RANK & The variable \textit{RANK} gives the rank of the submitted bids in tender $t$ from 1 (lowest bid) to 8 (highest bid). If there are more than 8 bids in a tender, all bids ranked above 8 take the value 8. \\
LNCAP & The natural logarithm of the backlog of current contracts for firm $i$ in tender $t$. \\
LNCAP\_SEC & The natural logarithm of the minimal backlog of current contracts held by rivals of firm $i$ in tender $t$. \\
LNDIST & The natural logarithm of the distance for firm $i$ in tender $t$. \\
LNDIST\_SEC & The natural logarithm of the minimal distance among rivals of firm $i$ in tender $t$. \\
LASTWIN & The variable \textit{LASTWIN} gives the number of days without winning for firm $i$ in tender $t$. \\
EQUAL & The variable \textit{EQUAL} gives the turnover already won by firm $i$ in tender $t$. \\
SCORE & The variable \textit{SCORE} gives the number of cover bids submitted by firm $i$ in tender $t$ since the last contract won. \\
SHAREH1 & The dummy variable \textit{SHAREH1} takes the value 1 if the firm is a shareholder in firm 1. \\
SHAREH2 & The dummy variable \textit{SHAREH2} takes the value 1 if the firm is a shareholder in firm 2. \\
VERTINT & The dummy variable \textit{VERTINT} takes the value 1 if the firm has its own mixing plant. \\
DUMMIES FOR FIRMS & Firm-specific dummy variables capturing individual characteristics of each firm. \\
\hline
\end{tabular}
\end{center}
\caption{Description of the variables} \label{desvarch1}
\par
\end{table}

\newpage

\section{The cartel convention at work}

\subsection{Analysis of the lowest bid}
We begin by examining whether contract allocation within the cartel followed a cost-based logic consistent with the convention. To do so, we estimate a series of logit models in which the dependent variable equals one if a bid is the lowest in a given tender. Tables \ref{tab:logit_models_indiv} and \ref{tab:logit_models_full} report the results for individual bids and for the full sample including consortium bids, respectively. The different specifications progressively introduce cost proxies, allocation variables, market structure controls, and firm fixed effects.

Across all specifications, the proxy for distance to the contract site (LNDIST) is highly significant and negative. In Table \ref{tab:logit_models_indiv}, the coefficient ranges between $-0.73$ and $-0.92$, while in Table \ref{tab:logit_models_full} it ranges between $-0.54$ and $-0.98$, always significant at the 1\% level. A greater distance between a firm’s headquarters and the project location therefore substantially reduces the probability of submitting the lowest bid. This result reflects standard economic logic: higher transport and logistical costs translate into higher bids and lower competitiveness. More importantly, it is fully consistent with the cartel’s own allocation rules, which explicitly prioritized distance as a central cost criterion in the convention.

The variable capturing the minimum distance among competitors (LNDIST\_SEC) is also robustly significant and positive. In Table \ref{tab:logit_models_indiv}, its coefficient lies between $0.83$ and $0.99$, and in Table \ref{tab:logit_models_full} between $0.53$ and $1.10$. This implies that the probability of winning increases when the closest rival is farther away from the project site. In cartel terms, designated losers typically have higher distances than the designated winner, reinforcing the interpretation that the cartel internally ranked firms according to relative cost positions rather than allocating contracts arbitrarily.

Turning to capacity constraints, measured by the backlog of ongoing contracts (LNCAP), we observe an important difference between individual bids and the full sample. In Table \ref{tab:logit_models_indiv}, which uses only individual bids, the coefficients on LNCAP and LNCAP\_SEC are small and statistically insignificant. By contrast, once consortium bids are included (Table \ref{tab:logit_models_full}), LNCAP becomes negative and highly significant across all specifications, with coefficients between $-0.37$ and $-0.42$. A higher backlog therefore reduces the probability of submitting the lowest bid, which again follows a clear cost logic: firms already committed to many projects face higher marginal costs and thus bid less aggressively.

Two mechanisms explain why capacity becomes significant when consortium bids are included. First, the number of observations increases from 880 to 1,325, improving statistical precision. Second, consortium members share project costs, which mechanically reduces within-group variance and sharpens the estimation of capacity effects. More substantively, the convention itself helps explain this pattern: for contracts above CHF 500,000, allocation was based not only on distance but also on firms’ workload. Since consortia are mainly formed for larger contracts, including them effectively increases the weight of backlog considerations in the allocation mechanism, exactly as the regressions reveal.

We next consider variables capturing explicit allocation rules. LASTWIN, measuring whether a firm recently won a contract, is weakly significant only in the most parsimonious specifications and becomes insignificant once cost variables are included. This suggests that intertemporal rotation played a limited role compared with cost-based allocation. The variable EQUAL, which sums contracts won during the cartel period, is positive and significant in both tables, with coefficients around $0.05$--$0.07$ in Table \ref{tab:logit_models_indiv} and $0.02$--$0.06$ in Table \ref{tab:logit_models_full}. Rather than reflecting a pure fairness-based equalization scheme, this coefficient captures a size effect: larger firms systematically win more contracts than smaller ones.

The variable SCORE, measuring contributions through cover bidding, delivers mixed evidence. In Table \ref{tab:logit_models_indiv}, SCORE is statistically insignificant throughout, although negative in the simpler specifications. This reflects the obligation imposed by the convention that all cartel members must participate in tenders. In the full sample (Table \ref{tab:logit_models_full}), however, SCORE becomes positive and significant in the most complete specification, with a coefficient of about $0.048$. This indicates that, once consortia and firm heterogeneity are controlled for, firms that contribute more actively to cartel discipline through cover bids enjoy a slightly higher probability of winning, consistent with implicit incentive mechanisms inside the cartel.

As SCORE is significant in Model 6 in Table \ref{tab:logit_models_full}, we compare both the relevance of cartel discipline (SCORE) and distance (LNDIST) for winning a contract. The coefficient on \textit{LNDIST} is large and negative ($-0.968$), implying an odds ratio of $\exp(-0.968)\approx 0.38$. Thus, a one-unit increase in log-distance reduces the odds of submitting the lowest bid by about $62\%$.
By contrast, the coefficient on \textit{SCORE} is much smaller ($0.048$), corresponding to an odds ratio of $\exp(0.048)\approx 1.05$. Sustaining the cartel one additional time therefore raises the odds of winning by only about $5\%$.
This comparison illustrates that cost-related factors dominate cartel allocation variables when they are significant: while participation in cover bidding marginally improves future winning chances, distance to the site plays a far more important role in determining who wins. The Ticino cartel therefore relied primarily on underlying cost proxies to allocate contracts.

Model comparison criteria further support the central role of costs. Models including only cost proxies outperform those relying exclusively on allocation variables, as shown by lower AIC and higher log-likelihood values for Model 1 relative to Model 2 in both tables (e.g. AIC $667$ vs.\ $696$ in Table \ref{tab:logit_models_indiv}, and $1269$ vs.\ $1332$ in Table \ref{tab:logit_models_full}). The best-performing specifications are Models 5 and 6, which include firm fixed effects. This indicates that, beyond cost proxies, persistent firm-level heterogeneity matters, but it does not overturn the main result: the cartel primarily allocated contracts according to observable cost characteristics.

As a robustness analysis, we evaluate whether observable cost proxies or contract allocation variables better explain the lowest bids in a tender with machine learning (see Appendix \ref{AppendixB}). Using lasso and ensemble learning methods, we predict whether a bidder submits the lowest bid when restricting attention to the first two ranks in each tender. Across specifications, models based on cost variables consistently outperform those relying on contract allocation rules. Ensemble methods achieve out-of-sample accuracy around 68--70\%, with balanced performance across winners and runners-up, whereas models using only allocation variables perform close to random classification. These findings again confirm that the designation of the winner by the cartel is primarily driven by cost-related factors rather than by formal allocation criteria alone.

Overall, the analyses of lowest bids provide strong evidence that the Ticino cartel used a ranking mechanism, closely mimicking competitive cost-based allocation. Rather than distributing contracts mechanically or politically, the cartel assigned wins to firms with the lowest implied costs, thereby approximating the first-best collusive outcome described in the theoretical literature on weak cartels.

\newpage

\begin{table}[h]
\begin{center}
\begin{tabular}{l D{.}{.}{4.6} D{.}{.}{4.6} D{.}{.}{4.6} D{.}{.}{4.6} D{.}{.}{4.6} D{.}{.}{4.6}}
\toprule
 & \multicolumn{1}{c}{Model 1} & \multicolumn{1}{c}{Model 2} & \multicolumn{1}{c}{Model 3} & \multicolumn{1}{c}{Model 4} & \multicolumn{1}{c}{Model 5} & \multicolumn{1}{c}{Model 6} \\
\midrule
(Intercept)    & -2.375^{***} & -2.563^{***} & -2.490^{***} & -1.901^{***} & -0.747       & -1.211       \\
               & (0.502)      & (0.175)      & (0.474)      & (0.570)      & (0.542)      & (1.174)      \\
LNCAP         & -0.018       &              & -0.113       & -0.112       & -0.164       & -0.143       \\
               & (0.201)      &              & (0.181)      & (0.185)      & (0.185)      & (0.185)      \\
LNDIST        & -0.764^{***} &              & -0.725^{***} & -0.800^{***} & -0.922^{***} & -0.924^{***} \\
               & (0.130)      &              & (0.130)      & (0.139)      & (0.164)      & (0.173)      \\
LNDIST\_SEC   & 0.987^{***}  &              & 0.933^{***}  & 0.986^{***}  & 0.833^{***}  & 0.849^{***}  \\
               & (0.267)      &              & (0.264)      & (0.262)      & (0.243)      & (0.248)      \\
LNCAP\_SEC    & 0.183        &              & 0.124        & 0.121        & 0.107        & 0.115        \\
               & (0.109)      &              & (0.099)      & (0.098)      & (0.093)      & (0.093)      \\
LASTWIN        &              & 0.003^{*}    & 0.002^{*}    & 0.002        & 0.000        & 0.000        \\
               &              & (0.001)      & (0.001)      & (0.001)      & (0.001)      & (0.001)      \\
SCORE         &              & -0.061       & -0.056       & -0.042       & 0.032        & 0.033        \\
               &              & (0.032)      & (0.032)      & (0.033)      & (0.035)      & (0.035)      \\
EQUAL   &              & 0.071^{***}  & 0.050^{***}  & 0.056^{***}  & 0.059^{***}  & 0.059^{***}  \\
               &              & (0.012)      & (0.012)      & (0.012)      & (0.014)      & (0.013)      \\
VERTINT        &              &              &              & -0.335       &              & 0.257        \\
               &              &              &              & (0.479)      &              & (1.130)      \\
SHAREH\_A        &              &              &              & -0.717       &              & 0.393        \\
               &              &              &              & (0.407)      &              & (1.122)      \\
SHAREH\_B        &              &              &              & -0.055       &              & 0.756        \\
               &              &              &              & (0.412)      &              & (1.191)      \\
\midrule
Dummy firms &  0       &  0    &       0       &   0           & 12      & 12     \\
AIC            & 667.291      & 696.181      & 664.787      & 667.439      & 646.853      & 652.145      \\
BIC            & 691.190      & 715.301      & 703.027      & 720.018      & 742.451      & 762.084      \\
Log Likelihood & -328.645     & -344.091     & -324.394     & -322.720     & -303.426     & -303.073     \\
Deviance       & 657.291      & 688.181      & 648.787      & 645.439      & 606.853      & 606.145      \\
Num. obs.      & 880          & 880          & 880          & 880          & 880          & 880          \\
\bottomrule
\multicolumn{7}{l}{\scriptsize{$^{***}p<0.001$; $^{**}p<0.01$; $^{*}p<0.05$}}
\end{tabular}
\caption{Binary logit models for individual bids (outcome: lowest bid). Standard errors clustered at the auction (project) level.}
\label{tab:logit_models_indiv}
\end{center}
\end{table}

\begin{table}[h]
\begin{center}
\begin{tabular}{l D{.}{.}{4.6} D{.}{.}{4.6} D{.}{.}{4.6} D{.}{.}{4.6} D{.}{.}{4.6} D{.}{.}{4.6}}
\toprule
 & \multicolumn{1}{c}{Model 1} & \multicolumn{1}{c}{Model 2} & \multicolumn{1}{c}{Model 3} & \multicolumn{1}{c}{Model 4} & \multicolumn{1}{c}{Model 5} & \multicolumn{1}{c}{Model 6} \\
\midrule
(Intercept)    & -1.592^{***} & -1.697^{***} & -1.590^{***} & -0.988^{**}  & -0.090       & 0.649        \\
               & (0.291)      & (0.114)      & (0.283)      & (0.359)      & (0.359)      & (0.410)      \\
LNCAP         & -0.366^{***} &              & -0.421^{***} & -0.413^{***} & -0.403^{***} & -0.397^{**}  \\
               & (0.105)      &              & (0.111)      & (0.115)      & (0.122)      & (0.126)      \\
LNDIST        & -0.548^{***} &              & -0.543^{***} & -0.575^{***} & -0.979^{***} & -0.968^{***} \\
               & (0.143)      &              & (0.146)      & (0.150)      & (0.187)      & (0.185)      \\
LNDIST\_SEC   & 1.103^{***}  &              & 1.038^{***}  & 1.083^{***}  & 0.534^{*}    & 0.593^{**}   \\
               & (0.228)      &              & (0.226)      & (0.231)      & (0.212)      & (0.221)      \\
LNCAP\_SEC    & 0.265^{**}   &              & 0.229^{**}   & 0.224^{**}   & 0.213^{*}    & 0.211^{*}    \\
               & (0.084)      &              & (0.083)      & (0.083)      & (0.098)      & (0.097)      \\
LASTWIN        &              & 0.002^{***}  & 0.002^{**}   & 0.001        & 0.000        & -0.000       \\
               &              & (0.001)      & (0.001)      & (0.001)      & (0.001)      & (0.001)      \\
SCORE         &              & -0.034       & -0.025       & -0.009       & 0.027        & 0.048^{*}    \\
               &              & (0.019)      & (0.019)      & (0.020)      & (0.022)      & (0.023)      \\
EQUAL   &              & 0.023^{*}    & 0.023^{*}    & 0.029^{**}   & 0.056^{***}  & 0.060^{***}  \\
               &              & (0.009)      & (0.011)      & (0.011)      & (0.011)      & (0.011)      \\
VERTINT        &              &              &              & -0.269       &              & -0.412       \\
               &              &              &              & (0.271)      &              & (0.280)      \\
SHAREH\_A        &              &              &              & -0.738^{*}   &              & -1.000^{**}  \\
               &              &              &              & (0.287)      &              & (0.356)      \\
SHAREH\_B        &              &              &              & -0.369       &              & -0.367       \\
               &              &              &              & (0.288)      &              & (0.368)      \\
\midrule
Dummy firms &  0       &  0    &       0       &   0           & 12      & 12     \\
AIC            & 1268.964     & 1332.165     & 1267.790     & 1266.326     & 1189.428     & 1186.298     \\
BIC            & 1294.909     & 1352.922     & 1309.303     & 1323.407     & 1293.212     & 1305.649     \\
Log Likelihood & -629.482     & -662.082     & -625.895     & -622.163     & -574.714     & -570.149     \\
Deviance       & 1258.964     & 1324.165     & 1251.790     & 1244.326     & 1149.428     & 1140.298     \\
Num. obs.      & 1325         & 1325         & 1325         & 1325         & 1325         & 1325         \\
\bottomrule
\multicolumn{7}{l}{\scriptsize{$^{***}p<0.001$; $^{**}p<0.01$; $^{*}p<0.05$}}
\end{tabular}
\caption{Binary logit models for individual bids (outcome: lowest bid). Standard errors clustered at the auction (project) level.}
\label{tab:logit_models_full}
\end{center}
\end{table}

\newpage

\subsection{Analysis of the ranks of bids}
We complement the binary analysis of winning bids by examining the full ordering of bids within each tender. Rather than focusing only on who wins, we analyze how the cartel structured the entire distribution of bids. We estimate ordered logit models in which the dependent variable is the rank of each bid, with higher values corresponding to less competitive offers. Table \ref{ordlogit1} reports results for individual bids only, while Table \ref{ordlogit2} extends the sample to include consortia.\footnote{The proportional-odds assumption underlying the ordered logit models was evaluated using a formal diagnostic test. For Models 1 to 3, the test failed to reject the null hypothesis that the proportional-odds assumption holds. For Model 4, the test statistic could not be computed because the information matrix was singular and therefore not invertible, likely reflecting the increased model complexity and limited number of observations. As a result, the validity of the proportional-odds assumption cannot be formally assessed for Model 4.}

The rank regressions strongly confirm and extend the lowest-bid results. In both tables, the distance variable (LNDIST) is positive and highly significant across all specifications. In Table \ref{ordlogit1}, the coefficient ranges from about $0.57$ to $0.71$, while in Table \ref{ordlogit2} it ranges from about $0.42$ to $0.79$. This implies that firms located farther from the project site systematically submit bids that are ranked worse. Distance therefore captures economically meaningful cost variation, and the cartel clearly incorporated this ordering when assigning bid positions. Rather than randomizing ranks, the cartel aligned bid placement with observable cost proxies, preserving a competitive-looking structure.

The same pattern emerges for rival distance. The coefficient on LNDIST\_SEC is negative and statistically significant in both samples. In Table \ref{ordlogit1}, estimates lie around $-0.25$ to $-0.35$, while in Table \ref{ordlogit2} they are larger in magnitude, between about $-0.38$ and $-0.79$. When the closest rival has a stronger geographic cost advantage, a firm is more likely to appear at a worse rank. In cartel terms, losing bidders typically face a designated winner with substantially lower distance-based costs, revealing that the cartel sorted firms internally according to cost before bids were submitted.

Capacity effects are more nuanced. For individual bids (Table \ref{ordlogit1}), LNCAP and LNCAP\_SEC are small and insignificant (e.g.\ LNCAP around $-0.16$ to $-0.19$). In the full sample (Table \ref{ordlogit2}), LNCAP becomes positive and significant in the parsimonious models (about $0.25$ in Models 3--4), indicating that firms with heavier backlogs submit worse-ranked bids. Likewise, LNCAP\_SEC is negative and significant (around $-0.19$ to $-0.20$ in Models 3--4), suggesting that losing bidders tend to be more capacity-constrained than winners. Once firm fixed effects are introduced (Models 5--6), these effects weaken, implying that backlog matters, but much of its influence reflects persistent firm characteristics rather than short-term tactical allocation.

Turning to allocation variables, LASTWIN plays only a limited role. Its coefficient is close to zero in Table \ref{ordlogit1} and only weakly significant in the full sample (about $-0.001$ to $-0.002$ in Table \ref{ordlogit2}), confirming that short-run rotation rules were secondary in determining bid ranks. By contrast, EQUAL is consistently negative and significant in both tables. In Table \ref{ordlogit1}, coefficients range from about $-0.06$ to $-0.07$, and in Table \ref{ordlogit2} from about $-0.03$ to $-0.07$. Firms that have accumulated more contracts tend to submit more competitive bids and appear at better ranks. This reflects differences in size and operational capacity rather than a strict equal-sharing rule imposed by the cartel.

The variable SCORE is particularly informative for cartel discipline. In the individual-bid sample (Table \ref{ordlogit1}), SCORE is small and insignificant. In the full sample (Table \ref{ordlogit2}), however, it is positive and significant in simpler specifications (around $0.03$--$0.05$ in Models 3--4). This implies that firms providing more cover bids tend to appear at worse ranks. This captures a core feature of cartel organization: members are required to submit deliberately non-competitive offers to support the designated winner, mechanically pushing their bids toward superior positions. Once firm fixed effects are included, the effect vanishes, indicating that cover-bidding behavior is closely tied to stable firm roles within the cartel.

Model fit statistics further support these conclusions. In both tables, specifications including firm dummies (Models 5--6) substantially improve log-likelihood and AIC values. Importantly, however, the qualitative message is unchanged across models: bid ranks are primarily shaped by cost-based proxies embedded in the convention rather than by arbitrary redistribution mechanisms.

Taken together, the rank analysis strengthens the paper's central and policy-relevant claim. The Ticino cartel did not merely manipulate winners' bids; it organized the entire distribution of bids to mirror underlying cost structures. By aligning both winning probabilities and bid rankings with distance and capacity, the cartel generated outcomes that look competitive while preserving collusive efficiency. This helps explain why standard econometric screening tools based on cost patterns might fail to detect sophisticated procurement cartels: the statistical signature of competition is deliberately reproduced inside the collusive process itself.

\begin{table}[!htp]
\begin{center}
\begin{tabular}{l D{.}{.}{5.6} D{.}{.}{5.6} D{.}{.}{5.6} D{.}{.}{5.6}}
\toprule
 & \multicolumn{1}{c}{Model 3} & \multicolumn{1}{c}{Model 4} & \multicolumn{1}{c}{Model 5} & \multicolumn{1}{c}{Model 6} \\
\midrule
LNCAP         & -0.172       & -0.160       & -0.192       & -0.194       \\
               & (0.137)      & (0.129)      & (0.138)      & (0.139)      \\
LNDIST        & 0.566^{***}  & 0.647^{***}  & 0.708^{***}  & 0.698^{***}  \\
               & (0.144)      & (0.141)      & (0.136)      & (0.139)      \\
LNDIST\_SEC   & -0.246^{*}   & -0.314^{**}  & -0.349^{**}  & -0.331^{**}  \\
               & (0.113)      & (0.111)      & (0.117)      & (0.121)      \\
LNCAP\_SEC    & -0.127       & -0.117       & -0.094       & -0.088       \\
               & (0.068)      & (0.068)      & (0.070)      & (0.069)      \\
LASTWIN        & -0.001       & -0.000       & 0.000        & 0.000        \\
               & (0.001)      & (0.001)      & (0.001)      & (0.001)      \\
SCORE         & 0.022        & 0.005        & -0.025       & -0.021       \\
               & (0.020)      & (0.021)      & (0.022)      & (0.021)      \\
EQUAL   & -0.056^{***} & -0.066^{***} & -0.063^{***} & -0.061^{***} \\
               & (0.012)      & (0.012)      & (0.012)      & (0.011)      \\
VERTINT        &              & -0.693^{*}   &              & -1.132       \\
               &              & (0.312)      &              & (1.010)      \\
SHAREH\_A        &              & 0.433        &              & -1.583       \\
               &              & (0.310)      &              & (1.007)      \\
SHAREH\_B       &              & 0.463        &              & -0.762       \\
               &              & (0.248)      &              & (1.253)      \\
\midrule
AIC            & 3563.070     & 3534.034     & 3489.949     & 3492.741     \\
BIC            & 3629.989     & 3615.293     & 3614.227     & 3631.358     \\
Log Likelihood & -1767.535    & -1750.017    & -1718.974    & -1717.370    \\
Deviance       & 3535.070     & 3500.034     & 3437.949     & 3434.741     \\
Num. obs.      & 880          & 880          & 880          & 880          \\
\bottomrule
\multicolumn{5}{l}{\scriptsize{$^{***}p<0.001$; $^{**}p<0.01$; $^{*}p<0.05$}}
\end{tabular}
\caption{Ordered logit models for individual bids (outcome: bid rank). Standard errors clustered at the auction (project) level.}
\label{ordlogit1}
\end{center}
\end{table}

\begin{table}[!htp]
\begin{center}
\begin{tabular}{l D{.}{.}{5.6} D{.}{.}{5.6} D{.}{.}{5.6} D{.}{.}{5.6}}
\toprule
 & \multicolumn{1}{c}{Model 3} & \multicolumn{1}{c}{Model 4} & \multicolumn{1}{c}{Model 5} & \multicolumn{1}{c}{Model 6} \\
\midrule
LNCAP        & 0.254^{*}    & 0.257^{*}    & 0.170        & 0.165        \\
               & (0.128)      & (0.129)      & (0.102)      & (0.104)      \\
LNDIST        & 0.418^{*}    & 0.433^{**}   & 0.787^{***}  & 0.779^{***}  \\
               & (0.164)      & (0.160)      & (0.136)      & (0.135)      \\
LNDIST\_SEC   & -0.735^{***} & -0.792^{***} & -0.384^{***} & -0.403^{***} \\
               & (0.120)      & (0.119)      & (0.109)      & (0.108)      \\
LNCAP\_SEC    & -0.197^{**}  & -0.192^{**}  & -0.127^{*}   & -0.128^{*}   \\
               & (0.074)      & (0.074)      & (0.064)      & (0.064)      \\
LASTWIN        & -0.002^{***} & -0.001^{*}   & -0.001       & -0.000       \\
               & (0.000)      & (0.001)      & (0.000)      & (0.001)      \\
SCORE         & 0.045^{**}   & 0.034^{*}    & -0.001       & -0.010       \\
               & (0.014)      & (0.014)      & (0.017)      & (0.017)      \\
EQUAL   & -0.031^{*}   & -0.035^{**}  & -0.066^{***} & -0.067^{***} \\
               & (0.012)      & (0.013)      & (0.010)      & (0.010)      \\
VERTINT       &              & -0.304       &              & 0.223        \\
               &              & (0.241)      &              & (0.294)      \\
SHAREH\_A        &              & 0.385        &              & 0.655^{*}    \\
               &              & (0.256)      &              & (0.327)      \\
SHAREH\_B        &              & 0.359^{*}    &              & 0.422        \\
               &              & (0.182)      &              & (0.323)      \\
\midrule
AIC            & 5148.725     & 5135.625     & 4891.893     & 4891.990     \\
BIC            & 5221.374     & 5223.841     & 5026.812     & 5042.476     \\
Log Likelihood & -2560.363    & -2550.812    & -2419.947    & -2416.995    \\
Deviance       & 5120.725     & 5101.625     & 4839.893     & 4833.990     \\
Num. obs.      & 1325         & 1325         & 1325         & 1325         \\
\bottomrule
\multicolumn{5}{l}{\scriptsize{$^{***}p<0.001$; $^{**}p<0.01$; $^{*}p<0.05$}}
\end{tabular}
\caption{Ordered logit models for individual bids (outcome: bid rank). Standard errors clustered at the auction (project) level.}
\label{ordlogit2}
\end{center}
\end{table}

\newpage

\subsection{Cost-based detection method}
To assess whether the observed ranking of bids reflects underlying cost differences across firms, we apply the cost-based detection method proposed by \citet{Bajari2003}. The core idea is simple: in a competitive environment, firms’ bids should be ordered according to their costs, so that conditional on observable cost proxies, bids are independent across firms and exchangeable. If firms truly classify and rank their bids based on cost, these conditions should not be rejected. In contrast, if bid ranks are driven by coordination rather than costs, the ordering of bids will deviate from cost-based logic, generating correlation across bidders and heterogeneous responses to cost proxies. The framework of \citet{Bajari2003} therefore provides a natural benchmark to test whether bidding behavior is consistent with competitive, cost-based ranking or instead reflects coordinated distortions of bid orderings.

As shown in \citet{Bajari2003}, it is often more practical to use a regression approach rather than nonparametrically recovering costs from bids using auction models. The so-called reduced-form bidding function can be estimated using proxies for firms' own costs and rivals' costs as explanatory variables, allowing one to test conditional independence and exchangeability of bids. Most empirical studies adopt this approach in panel settings to analyze the structural relationship between cost proxies and bids \citep[see, e.g.,][]{Porter1993, Porter1999, Pesendorfer2000, Bajari2003, Ishii2009, Chotibhongs2012a, Chotibhongs2012b, Aryal2013, Lundberg2025}.

Appendix \ref{AppendixC} documents the econometric testing procedures and presents the complete set of raw results for the cartel and post-cartel periods. Table \ref{bajari} summarizes the Bajari and Ye (2003) test results using firm-specific coefficients. During the cartel period, only 17 out of 103 firm pairs fail the conditional independence test, implying that about 83\% of pairs go undetected.\footnote{The rejection frequencies reported in the summary tables are based on unadjusted p-values. While Benjamini–Hochberg (BH) adjusted p-values were also computed and are reported in Appendix, the main discussion emphasizes unadjusted p-values because the primary objective is to evaluate the tests' ability to detect potential cartel behavior. Multiple-testing corrections are designed to control false discoveries but can substantially reduce statistical power, increasing the risk of false negatives. For transparency, both adjusted and unadjusted results are available, although the summary statistics are based on the latter.} The exchangeability test flags 34 pairs (around 33\%), meaning that 67\% pass as competitive. For the post-cartel period, results must be interpreted cautiously due to the much smaller sample size. The conditional independence test flags 4 out of 47 pairs (about 8\%), while the exchangeability test flags 14 pairs (54\%), implying that 46\% pass as competitive.

Table \ref{tab:pairwise_tests_combined_periods} combines both tests at the pair level. During the cartel period, only 4\% of firm pairs are rejected by both tests simultaneously, while 42\% are not rejected by either test. The majority of pairs (54\%) are rejected by only one of the two tests. In the post-cartel period, 8\% of pairs are rejected by both tests, whereas 52\% are not rejected by either. These findings reinforce the broader conclusion that standard econometric screening methods based on conditional independence and exchangeability have limited power to detect sophisticated bid-rigging arrangements when collusive bids are organized according to underlying cost structures.

{\renewcommand{\arraystretch}{1.1}
\begin{table}[!htp]
\centering
\begin{tabular}{llrrrrrrrr}
\hline
& & \multicolumn{4}{c}{Conditional Independence} 
& \multicolumn{4}{c}{Exchangeability} \\
\cline{3-6} \cline{7-10}
N.pairs & Period 
& True & Perc. & False & Perc. 
& True & Perc.& False & Perc. \\
\hline
103 & Cartel 
& 17 & 17\% & 86 & 83\% 
& 34 & 33\% & 69 & 67\% \\

47 (26) & Post-Cartel 
& 4 & 8\% & 43 & 92\% 
& 14 & 54\% & 12 & 46\% \\
\hline
\end{tabular}
\caption{Synthesis of the Econometric Tests}
\label{bajari}
\vspace{0.2cm}
{\footnotesize
\textit{Note:} ``Conditional Independence'' and ``Exchangeability'' refer to the Bajari--Ye conditional independence test and the exchangeability test, respectively. Percentages are computed relative to the total number of firm pairs in each period.
}
\end{table}
}

\begin{table}[ht]
\centering
\begin{tabular}{lrrrr}
\hline
& \multicolumn{2}{c}{Cartel Period} & \multicolumn{2}{c}{Post-Cartel Period} \\
\cline{2-3} \cline{4-5}
Status & N.pairs & Perc. & N.pairs & Perc. \\
\hline
Both & 4  & 4\%  & 2  & 8\%  \\
One  & 56 & 54\% & 11 & 42\% \\
None & 43 & 42\% & 13 & 52\% \\
\hline
\end{tabular}
\caption{Combined Rejection Summary by Pair Across Periods}
\label{tab:pairwise_tests_combined_periods}
\end{table}

Setting aside the potential risk of false positives in the post-cartel period, the rate of false negatives during the cartel period is substantial. Rather than undermining econometric screening tools per se, this pattern reflects the particular sophistication of the Ticino cartel. As emphasized by \citet{Bajari2003}, collusion may remain undetected when firms inflate bids in ways that remain aligned with underlying cost structures. In Ticino, the convention explicitly required contract allocation and bid coordination to follow cost-based rules using pre-calculated bids. As a result, cartel members ranked bids consistently with cost proxies and assigned cover bids accordingly. The large number of undetected firm pairs therefore provides additional evidence that allocation was genuinely cost-driven and that the convention was closely adhered to in practice. The robustness exercises therefore highlight an important limitation of cost-based methods: sophisticated cartels that organize around efficiency principles evading detection.

\section{Estimating the cartel damage}

This section presents an empirical analysis of the economic damage caused by the Ticino cartel, measured as the average price differential between the collusive and competitive periods after adjusting for control variables. Since the cartel period spanned from 1999 up to and including March 2005, followed by a competitive period from April 2005 to April 2006, all bids are deflated to 1999 real prices to account for inflation.

The literature provides several studies on the estimation of cartel damage. For instance, \citet{kwoka1997} analyzes a long-running bidding conspiracy in real estate auctions in Washington, DC, and finds estimated damages between 22\% and 32\%, reflecting bidders’ incentives to shade their bids in knockout auctions. \citet{connor2014} offers a meta-analysis of nearly 600 studies and judicial decisions, comprising 1,517 quantitative estimates of cartel overcharges. The median overcharge across all cartels is 23.3\%, with international cartels showing higher overcharges (30.0\%) than domestic ones (17.2\%). Due to the skewed distribution of overcharges, the mean is even higher at 50.4\%. The study also argues that assuming a 10\% overcharge for penalty calculations is inadequate and calls for stronger deterrence.

\citet{GABRIELLI2023106404} introduce a model-based method for estimating damages from bidding rings in first-price auctions, using cartelized auction data to recover bid distributions of both cartel and non-cartel participants. These distributions are then used to infer counterfactual competitive prices and estimate overcharges. In contrast, our approach requires fewer structural assumptions but relies on data from both competitive and collusive tenders, in which all bidders were part of the cartel. Similarly, \citet{caoui2022} use both collusive and competitive data in a structural analysis of the Texas school milk cartel and additionally estimate umbrella damages caused by non-cartel firms raising prices when bidding alongside the cartel. They find that such indirect damages account for at least 35\% of total harm. Our study differs from these and other contributions (e.g. \citealp{nieberding2006}) in that, to the best of our knowledge, it is the first to use flexible nonparametric methods to estimate cartel damage. Nonetheless, our findings are consistent with the broader literature documenting substantial overcharges from cartel activity.

\subsection{Methods}

To assess the impact of cartel activity on bidding behavior, we apply the double machine learning (DML) approach of \cite{Chetal2018}, using the logarithm of bids (\textit{LNBID}) as the dependent variable. DML estimates the average treatment effect by comparing bids submitted during cartel and competitive periods while controlling for covariates in a data-driven way. These covariates include cost proxies: the logarithm of a firm’s backlog (\textit{LNCAP}), the logarithm of distance to the contract location (\textit{LNDIST\textsubscript{it}}), and the logarithms of the minimum backlog and minimum distance among rival firms in a tender (\textit{LNCAP\_SEC}, \textit{LNDIST\_SEC}). 

For consortia, we additionally control for the minimum and mean values of these cost proxies across all consortium members. Further controls include firm fixed effects (with recurring consortia treated as distinct firms), seasonal fixed effects (quarterly dummies), and a dummy indicating whether the bid was submitted by a consortium or an individual firm.

We implement the method using the \textit{DoubleML} package for \textsf{R} \citep{DoubleML}, with 5-fold cross-fitting to mitigate overfitting. We repeat the random sample splitting and estimation 20 times and report the median estimate and standard error to reduce variance in effect estimation. For the nuisance components in the DML procedure, we use random forests \citep{Breiman2001} with 1,000 trees via the \textit{ranger} package \citep{ranger}.

To validate our identifying assumption--namely that after conditioning on covariates, bids from cartel and competitive periods are comparable--we conduct placebo tests using data from 1999 to 2004. We define years 2002--2004 as pseudo-treated ($D=1$) and 1999--2001 as pseudo-controls ($D=0$), although all belong to the cartel regime. The true placebo effect is therefore zero, and significant estimates would indicate violations of the identifying assumption. We perform placebo tests using either all bids in a tender or only the lowest bids, which is most relevant for procurement cost. Standard errors are clustered at the tender level when using all bids and at the firm level when using lowest bids.

\subsection{Results}

The first row in Table \ref{tab:plac} (1999--2004) reports the placebo results. For both all bids (right panel) and lowest bids (left panel), placebo effects are statistically insignificant and close to zero. As a robustness check, we restrict the sample to 2000--2003 and obtain similarly insignificant placebo estimates. We also conduct placebo tests for later cartel periods (2001--2004), treating 2001--2002 as pseudo-controls and 2003--2004 as pseudo-treated. These pseudo-effects remain small and far from conventional significance levels.

\begin{table}[htbp]  
	\begin{center}

		\begin{tabular}{c|cccc|cccc}
			\hline
			&\multicolumn{4}{c|}{ all bids } &\multicolumn{4}{c}{ lowest bids } \\
			years  & obs  & estimate & se  & p-value & obs  & estimate & se  & p-value\\
			\hline
			&\multicolumn{8}{c}{ Pseudo controls: year$<$ 2002 } \\
			1999-2004 & 983 & -0.0221 &  0.1182 & 0.8520 & 173 &  0.0692 &  0.1813 & 0.7030 \\
			2000-2003 & 558 &  0.0371 &  0.1874 & 0.8430 & 101 &  -0.0135 &  0.2631 &  0.9592 \\
				\hline
			&\multicolumn{8}{c}{ Pseudo controls: year$<$ 2003 } \\
			2001-2004 &   632 & -0.0738 & 0.1463 & 0.6137 &  118 & 0.1405 & 0.3479 & 0.6860 \\
			\hline
		\end{tabular}
	\end{center}
			\caption{Placebo tests based on DML}
		\label{tab:plac}
	\par
	{\scriptsize Notes: obs, estimate, se and p-value denote the number of observations, the point estimate of the cartel effect, the corresponding standard error and the p-value of the two-sided test of the null hypothesis that the coefficient is equal to zero. Standard errors are clustered at the tender level for all bids and at the firm level for lowest bids.}
\end{table}

We next estimate actual price effects of cartel versus competition, reported in Table \ref{tab:eff}. We first apply DML to the period 2003--2006, using 2003 to March 2005 as cartel ($D=1$) and the remainder as competitive ($D=0$). For all bids and lowest bids, the effects are highly significant, amounting to 0.630 and 0.478 log points, respectively. These imply price increases of roughly 61\% to 88\%.\footnote{For instance, $100\cdot(\exp(0.6302)-1)=87.81$.} Extending the sample to 2001--2006 yields similarly large effects between about 0.65 and 0.67 log points. The same holds when using the full period 1999--2006.

Finally, we focus on years close to the regime shift by applying DML only to 2004--2006. This mimics the intuition of a regression discontinuity design \citep{Thistlethwaite60,HahnToodKlaauw01}, with time as the running variable. Periods just before and after cartel dissolution should be more comparable in terms of business conditions. Although we lack sufficient observations for a formal RDD, restricting attention to adjacent years allows us to approximate its logic. Estimates in the last row of Table \ref{tab:eff} amount to 0.586 and 0.447 log points for all bids and lowest bids, implying price increases of roughly 56\% to 80\%.

	\begin{table}[htbp]
	\begin{center}
	
		\begin{tabular}{c|cccc|cccc}
			\hline
			&\multicolumn{4}{c|}{ all bids } &\multicolumn{4}{c}{ lowest bids } \\
			years  & obs  & estimate & se  & p-value & obs  & estimate & se  & p-value\\
				\hline	
			2003-2006 & 607 & 0.6302 &  0.1601  & 0.0000 & 113 & 0.4776 & 0.1507 & 0.0015 \\
			2001-2006 & 899 & 0.6668 &  0.1200  & 0.0000 & 166 & 0.6532 & 0.1528 &  0.0000 \\
			1999-2006 & 1250 & 0.6602 & 0.1060 & 0.0000 & 221 &  0.6027& 0.1382 & 0.0000 \\  
			2004-2006 & 507 &  0.5858  & 0.1844 &  0.0015 &  92 &  0.4470 & 0.1310 &  0.0006 \\
				\hline
		\end{tabular}
	\end{center}
		\caption{Effects based on DML}
		\label{tab:eff}
	\par
	{\scriptsize Notes: obs, estimate, se and p-value denote the number of observations, the point estimate of the cartel effect, the corresponding standard error and the p-value of the two-sided test of the null hypothesis that the coefficient is equal to zero. Standard errors are clustered at the tender level for all bids and at the firm level for lowest bids.}
\end{table}

As a further approach, we estimate a linear regression of log bids on a cartel dummy using the full sample (1999--2006), including firm and seasonal fixed effects, firm-level cost proxies, and minimum rival cost proxies. Consortium-level proxies are excluded to avoid multicollinearity. The specification also includes interactions between firm dummies and cost proxies, consistent with equation \eqref{eqestim2ch2}, and a linear time trend. While less flexible than DML, this approach allows explicit control for time trends. As shown in Table \ref{tab:eff2}, estimates are around 0.37 to 0.44 log points and statistically significant, implying price increases of roughly 45\% to 56\%.

To sum up, our estimated cartel damages of roughly 45\% to 90\% exceed the financial burdens typically reported in the literature. This again illustrates the exceptional nature of the Ticino cartel: although weak in appearance, it was able to fully implement the first-best collusive outcome.

	\begin{table}[htbp]
	\begin{center}

		\begin{tabular}{ccc|ccc}
			\hline
			\multicolumn{3}{c|}{ all bids (1250 obs)} &\multicolumn{3}{c}{ lowest bids (221 obs)} \\
			estimate & se  & p-value &  estimate & se  & p-value\\
			\hline
			   0.3748 & 0.1835 & 0.0442 &  0.444 & 0.1625 &  0.0179\\
			\hline
		\end{tabular}
	\end{center}
			\caption{Regression with linear trend (1999-2006)}
		\label{tab:eff2}
	\par
	{\scriptsize Notes: estimate, se and p-value denote the point estimate of the cartel effect, the corresponding standard error and the p-value of the two-sided test of the null hypothesis that the coefficient is equal to zero. Standard errors are clustered at the tender level for all bids and at the firm level for lowest bids.}
\end{table}

\newpage

\section{Policy recommendations}
The evidence from the Ticino bid-rigging cartel highlights a fundamental challenge for competition policy: sophisticated cartels can coordinate efficiently while deliberately mimicking competitive behavior, thereby evading cost-based detection methods and generating substantial economic harm. The findings of this study point to several important policy implications for improving both the detection and deterrence of collusion in public procurement.

First, detection strategies should not be restricted to cost-based detection methods. The Ticino case demonstrates that the cartel would not have been identified using standard cost-based tests such as those proposed by \citet{Bajari2003}, precisely because cartel members deliberately aligned their bids with observable cost proxies.
By contrast, simpler screening methods based on descriptive statistics of bid distributions, so-called “screens”, would likely have detected anomalies \citep[see][]{HuberImhof2019, Imhof2020}. This underscores a key lesson for law enforcement agencies: different detection methods are complementary and should be used jointly to minimize false negatives, that is, failures to detect collusion. Cost-based approaches should therefore be combined with methods focusing on irregularities in bid distributions, participation patterns, or ranking structures within tenders.

More broadly, competition authorities may initiate investigations through customer complaints, leniency applications, or whistleblower reports. Statistical screening methods constitute only one additional detection tool and, in practice, many procurement cartel investigations are triggered by leniency submissions or information provided by whistleblowers. Screening tools should thus be viewed as complements to these more “passive” detection channels. Importantly, screening methods, including machine learning techniques, can play a crucial supporting role in this context. They enable authorities to assess more objectively the scale and scope of alleged collusion, thereby helping to prioritize cases. This is particularly valuable because whistleblowers or leniency applicants often have only partial knowledge of the cartel, having participated in limited meetings or projects. While their testimony may be sufficient to trigger an investigation, it rarely provides a complete picture. Screening tools can therefore help reconstruct the broader cartel structure and assess the credibility of such reports. Taken together, these considerations argue for a multi-pronged detection framework that integrates diverse analytical methods with investigative mechanisms.

Second, procurement design and market structure must be carefully monitored, particularly with respect to strategic inputs and vertical relationships. While the Ticino cartel is exceptional in its level of organization, it reveals a broader and more generalizable insight: control over key inputs can play a central role in sustaining collusion. In this case, cross-shareholdings in upstream asphalt production created strong interdependencies among firms and enabled credible punishment mechanisms, such as exclusion from essential inputs, thereby reinforcing cartel discipline. It is important to note that such joint ownership structures are not inherently anti-competitive and may even generate efficiency gains. However, they represent a critical element in the overall assessment of market competition. When combined with other warning signs, such as unexpected price increases or persistently stable market shares, these structural features should prompt increased scrutiny by both procurement authorities and competition agencies. In parallel, procurement design should aim to reduce opportunities for coordination by encouraging entry, limiting excessive transparency among bidders, and carefully monitoring the use of consortia, which may otherwise serve as vehicles for collusion.

Third, the magnitude of the estimated overcharges in the Ticino case calls for both stronger deterrence and improved benchmarking practices in procurement. An important lesson concerns the role of procurement authorities themselves. In Ticino, engineers did not systematically conduct detailed analyses that would have revealed the substantial increase in prices over time. Instead, cost estimates were likely based on bids submitted in previous years. Such practices can quickly internalize cartel prices, effectively normalizing inflated price levels and masking the existence of collusion. This endogenous benchmarking creates a feedback loop in which cartel prices become the reference point for future contracts. To mitigate this risk, procurement authorities should also rely on independent cost estimates, external benchmarks, or cross-regional comparisons rather than relying solely on historical bid data.

Finally, leniency programs and whistleblower incentives should be reinforced, as insider information remains one of the most effective means of uncovering sophisticated cartels that are able to evade some statistical detection methods. Ensuring that leniency frameworks are transparent, predictable, and attractive, while also expanding protections and incentives for individual whistleblowers, is essential to destabilizing such arrangements. More broadly, competition policy should evolve toward a more proactive and risk-based approach, with particular attention to sectors characterized by repeated interactions, homogeneous products, and control over key inputs, such as construction and infrastructure. Regular monitoring of these high-risk markets, combined with the integration of quantitative screening tools and qualitative market intelligence, can significantly improve early detection and help prevent long-lasting economic harm.

In sum, the Ticino case demonstrates that some cartels can be both highly efficient and highly concealed. Addressing this challenge requires a comprehensive approach that combines multiple detection methods, strengthens institutional safeguards, and enhances both enforcement and procurement practices.

\section{Conclusion}
This study provides new evidence on the internal functioning, market effects, and detectability of bid-rigging cartels in public procurement through an unusually detailed analysis of the Ticino road-construction cartel in Switzerland. The exceptional documentary evidence available in this case allows us to reconstruct the cartel's internal governance and bidding logic with a level of precision that is uncommon in the literature. In particular, we show how the convention established a stable and highly coordinated allocation mechanism based on firms' relative cost characteristics, regular meetings, and repeated interactions, enabling cartel members to sustain collusion without relying on side payments.

Our findings indicate that the cartel implemented an allocation mechanism allowing the cartel to attain the first-best collusive outcome, as described in the theoretical literature. Contracts were systematically awarded to firms with lower production costs, while cover bids were structured to preserve the appearance of competitive tendering. Both traditional econometric specifications and more flexible machine-learning approaches consistently show that cost-related variables strongly predict bid rankings and winning outcomes. This suggests that the cartel's succeeded in reproducing internally efficient outcomes minimizing incentives to deviate from the agreement.

At the same time, the paper highlights an important limitation of cost-based cartel detection methods. Conventional econometric screens based on bid independence or exchangeability perform poorly in this setting and generate a high number of false negatives.
This is likely because the cartel designed its bidding behavior to mimic competitive outcomes while maintaining internal efficiency. By aligning bids with cost proxies and firm characteristics, the cartel obscured the usual statistical signals of collusion. For competition authorities, this implies that reliance on cost-based detection method alone may be insufficient in environments where firms internalize cost logic into their collusive strategies, and that detection frameworks should integrate complementary, data-driven, and institution-specific methods.

Finally, our damage estimates indicate that the cartel generated very large economic harm, with average overcharges of at least 45\% and potentially substantially higher depending on the specification. These estimates exceed central benchmarks in much of the existing cartel literature and suggest that efficient internal coordination can significantly amplify the welfare losses associated with collusion. The Ticino case therefore illustrates how sophisticated bid-rigging arrangements can simultaneously achieve internal stability, evade conventional detection tools, and impose severe costs on public budgets.

More broadly, the paper points to several promising directions for future research. A first avenue concerns the relationship between cartel governance structures and detectability, particularly the extent to which organizational sophistication affects the performance of empirical screening methods. A second avenue involves integrating structural economic models with modern machine-learning techniques to develop more robust and transferable detection frameworks. Finally, extending similar analyses to other sectors and institutional settings would help determine how generalizable cost-based collusive mechanisms are and which policy interventions are most effective in reducing both the incidence and the persistence of procurement cartels.

\newpage

\newpage
\setcounter{table}{0}
\renewcommand{\thetable}{A.\arabic{table}}

\appendix

\section{The Ticino convention and descriptive statistics}
\label{AppendixA}

\subsection{The Ticino convention}
{\linespread{0.9}\selectfont
\begin{center}
\textbf{\Large AGREEMENT}\footnote{See the convention of the Ticino bid-rigging cartel in decision Strassenbeläge Tessin (LPC 2008/1, pp. 78–80).}
\end{center}

\subsubsection*{1. Contracting Parties}

All paving companies registered in the Commercial Register and affiliated with ATIPS may participate in this Agreement.

\subsubsection*{2. Purpose}

The purpose of this Agreement is to guarantee quality in execution and an appropriate price level.  
The Agreement is based on the principle of mutual loyalty among the contracting parties.

\subsubsection*{3. Governing Body}

The decision-making body is the Assembly of Company Owners. It is chaired by a moderator appointed on a case-by-case basis.  

As a rule, decisions are taken by a majority of those present. In the event of disagreement, the decision shall be taken by secret ballot with a majority of those present (who are required to vote), with directly interested parties abstaining. If the number of attendees is even, the company with the highest quota shall abstain from voting.

\subsubsection*{4. Organization}

Companies must be represented by the owner or by a manager with signing authority registered in the Commercial Register.  

Unless otherwise decided by the Assembly, meetings are held weekly. During these meetings, contracts due until the end of the following week are discussed. Attendance is mandatory. In case of absence, companies must inform themselves of the matters discussed and strictly comply with the decisions taken.

Each participant in a tender must fully calculate its bid. Bids shall be discussed jointly in order to determine the appropriate level of the lowest bid, taking into account the interests of both the contracting authority and the companies.  

Cover bids must be prepared on the basis of justifiable cost calculations for each individual item. Identical cover bid amounts must be avoided.  

For public contracts, some bids must normally be submitted in such a way that they may be considered for award, ensuring that the contracting authority has a genuine possibility of choice among competing firms.

If a company waives its turn to perform a contract without valid reason, even if absent from the meeting, it must wait four weeks before again claiming work according to its quota.

\subsubsection*{5. Subject Matter of the Agreement}

The Agreement covers the following paving works:

\begin{enumerate}
\item[(a)] All works for public entities (including electricity and transport companies, etc.) awarded through public tender or private invitation to tender.
\item[(b)] All other works exceeding CHF 20,000.
\item[(c)] Other services for third parties related to paving works.
\end{enumerate}

\subsubsection*{6. Obligations}

Contracting companies must, as a rule, participate in tenders for works under Section 5, subject to the criteria set out in Section 7.  

Registration for public tenders and participation in site visits are mandatory.  

Signatory companies undertake not to supply third parties with products from mixing plants, rentals, labor, or other services related to paving works.  

Consortia for paving works may only be formed with companies that are signatories to this Agreement.  

Owners of mixing plants undertake to supply all mixed products exclusively to signatory companies at fair and competitive prices.

\subsubsection*{7. Allocation of Works}

Works are divided into two groups:

\begin{enumerate}
\item Works up to CHF 500,000 (including maintenance lots).
\item Works exceeding CHF 500,000 (including all motorway works, even if of lower value).
\end{enumerate}

Works in both groups may be executed in consortium at the discretion of the interested companies, after consultation with the Assembly.

Allocation shall be based on the following criteria:

\begin{enumerate}
\item[(a)] \textbf{Workload} (normally the predominant criterion, applied flexibly in the case of minor differences).  
For consortia, the relevant benchmark is the average quota of participating companies compared with other consortia. If not all companies form a consortium, the lowest quota among consortium members is compared with the individual quotas of other interested companies.
\item[(b)] \textbf{Regionality}, particularly for first-group works and local clients, provided that rotation quotas are not excessively penalizing.
\item[(c)] \textbf{Specialization}, where specific technical expertise is required.
\item[(d)] \textbf{Preliminary (study) bids}, especially in private works. A company that prepared the initial study bid shall be given preferential consideration, subject to the overall criteria.
\item[(e)] \textbf{Collegial discussions}, in order to resolve special cases without creating difficulties for contracting authorities or competing firms.
\end{enumerate}

\subsubsection*{8. Quotas}

Separate quotas are maintained for each company and for each work group, regardless of whether works are executed individually or in consortium.  

All works obtained with the support of at least one company are included in the rotation quota; others are recorded in a separate list (suspended works). Values are calculated inclusive of VAT.

The value of awarded contracts is counted as follows:

\begin{enumerate}
\item Up to CHF 1,000,000: 100\%
\item Next CHF 1,000,000: 90\%
\item Next CHF 1,000,000: 80\%
\item Next CHF 1,000,000: 60\%
\item Amounts exceeding this: 50\%
\end{enumerate}

The calculated amounts are added to the relevant quota, taking into account the individual factor established at the meeting of 15 December 1998.

Large multi-year projects may exceptionally be divided into annual portions, to be added to the quota at the beginning of each year, subject to Assembly approval.

If a contract is not awarded within two months of allocation, it is removed from the quota. If later awarded, it is immediately reinstated. Failure to report non-award results in the amount remaining in the quota for additional two-month periods.

Quotas are never reset, although they may be reduced by an equal amount for all companies. Suspended works are reviewed every two months.

\subsubsection*{9. Study Bids and Corporate Works}

Study bids for private works must be presented at calculation meetings and are added to the quota upon support.  

Study bids for public works must be announced in due time. The announcing company has priority, subject to the criteria of Section 7.

Works for companies (including sports centers) must be added to the quota at the start of works, unless directly supported. The amount (limited to sub-base and supply/laying of mix) must exceed CHF 50,000.

\subsubsection*{10. Mixed Works}

Natural and artificial stone works are counted at 50\% for the stone portion and 100\% for remaining paving works.

Mastic asphalt works in road construction are counted at 25\% for the mastic asphalt and waterproofing portion and 100\% for the remaining paving works, provided all competing companies give support.

\subsubsection*{11. Commitments}

Each company undertakes to cooperate loyally and discreetly to find reasonable solutions necessary for the implementation of this Agreement and to bear its proportional share of related costs.

\subsubsection*{12. Amendments}

Amendments to this Agreement may be decided only by the Assembly of Company Owners expressly convened for that purpose.

\subsubsection*{13. Final Provisions}

This Agreement enters into force upon signature and remains valid for an indefinite period. The notice period for termination is three months.

In any event, a departing company undertakes to support the remaining companies until all have reached the departing company’s quota level.

\vspace{1cm}

\noindent
Adopted on 15 December 1998.}

\newpage

\subsection{Descriptive statistics}

{\renewcommand{\arraystretch}{1.1}
\begin{table} [!htp]
\begin{center}
\begin{tabular}{rrrrrrrrr}
  \hline 
Firm & Win & Bid & Success & Integ. & Sh\_A & Sh\_B & Sales & Share \\ 
  \hline
3 & 13 & 43 & 0.3023 & 1 & 0 & 0 & 13.5 & 0.0544 \\ 
  4 & 15 & 119 & 0.1261 & 1 & 0 & 1 & 16.8 & 0.0675 \\ 
  5 & 19 & 33 & 0.5758 & 1 & 0 & 0 & 13.6 & 0.0547 \\ 
  6 & 7 & 51 & 0.1373 & 1 & 0 & 0 & 10.2 & 0.0412 \\ 
  7 & 4 & 10 & 0.4000 & 0 & 0 & 0 & 4.8 & 0.0193 \\ 
  8 & 19 & 95 & 0.2000 & 0 & 1 & 0 & 17.1 & 0.0689 \\ 
  9 & 20 & 160 & 0.1250 & 0 & 1 & 0 & 15.3 & 0.0618 \\ 
  10 & 16 & 104 & 0.1538 & 0 & 1 & 0 & 16.5 & 0.0665 \\ 
  11 & 16 & 66 & 0.2424 & 0 & 1 & 0 & 15.0 & 0.0604 \\ 
  12 & 29 & 46 & 0.6304 & 0 & 1 & 0 & 21.6 & 0.0869 \\ 
  14 & 17 & 90 & 0.1889 & 0 & 1 & 0 & 17.9 & 0.0719 \\ 
  15 & 24 & 148 & 0.1622 & 0 & 1 & 0 & 18.9 & 0.0762 \\ 
  16 & 11 & 29 & 0.3793 & 0 & 0 & 1 & 10.8 & 0.0437 \\ 
  17 & 17 & 107 & 0.1589 & 0 & 1 & 0 & 13.3 & 0.0536 \\ 
  18 & 16 & 63 & 0.2540 & 0 & 1 & 0 & 14.4 & 0.0582 \\ 
  19 & 15 & 105 & 0.1429 & 0 & 1 & 0 & 15.9 & 0.0639 \\ 
  20 & 11 & 56 & 0.1964 & 0 & 0 & 0 & 12.7 & 0.0510 \\ 
   \hline
\end{tabular}
\end{center}
\caption{Descriptive statistics per firm in the cartel period} \label{Tabledesc3}
\par
{\footnotesize Note:  Firm, Win, Bid, Success, Integ., Sh\_A, Sh\_B, Sales, Share denote the following variables: the firm identifier, the number of contracts won, the number of bids submitted, the success rate (calculated as the number of contracts won divided by the number of bids), a dummy variable equal to 1 if the firm is vertically integrated with its own mixing plant, a dummy variable indicating whether the firm is a shareholder of firm A, a dummy variable indicating whether the firm is a shareholder of firm B, the total revenue won (in million CHF), and the market share based on revenue won, respectively.}
\end{table}}

{\renewcommand{\arraystretch}{1.1}
\begin{table} [!htp]
\begin{center}
\begin{tabular}{lrrrrrrrr}
  \hline
Var\_Name & N & Mean & Sd & Min & Q.inf & Median & Q.sup & Max \\ 
  \hline
 LNCAP & 880 & 2.651 & 0.549 & 0.000 & 2.740 & 2.771 & 2.792 & 2.830 \\ 
 LNCAP\_SEC & 880 & 2.320 & 0.976 & 0.000 & 2.675 & 2.736 & 2.754 & 2.804 \\ 
 LNDIST & 880 & 1.157 & 0.525 & 0.000 & 1.044 & 1.395 & 1.492 & 1.722 \\ 
 LNDIST\_SEC & 8800 & 0.880 & 0.488 & 0.000 & 0.880 & 1.021 & 1.180 & 1.690 \\ 
 SCORE & 880 & 3.976 & 4.366 & 0.000 & 1.000 & 3.000 & 5.250 & 26.000 \\ 
 LASTWIN & 880 & 110.934 & 101.840 & 1.000 & 42.000 & 76.000 & 142.000 & 637.000 \\ 
 EQUAL & 880 & 8.750 & 5.271 & 0.000 & 4.274 & 8.589 & 13.272 & 21.203 \\ 
   \hline
\end{tabular}
\end{center}
\caption{Descriptive statistics for the logit estimation sample (individual bids, cartel period)} \label{Tabledesc5}
\par
\end{table}}

\newpage

\newpage
\setcounter{table}{0}
\renewcommand{\thetable}{B.\arabic{table}}
\section{Machine learning-based prediction}
\label{AppendixB}
To further assess the predictive performance of cost variables and contract allocation variables, we employ machine learning techniques and evaluate predictive accuracy, defined as the correct classification rate. Specifically, we use lasso logit regression \citep{Tibshirani96} and an ensemble method \citep[e.g.,][]{Zhou2012} that combines several algorithms through a weighted average to predict the probability of submitting the lowest bid based on the control variables used with the logit and ordered logit regression. We restrict the analysis to the first and second lowest bids in each tender in the full sample. This restriction ensures a balanced sample of winners and runners-up within each tender, which is crucial for learning a model that predicts the likelihood of submitting the lowest bid with comparable accuracy across classes.

We use the \textit{SuperLearner} package by \citet{PolleyLeDellVanderLaan2016} for \textsf{R} to implement ensemble learning and the \textit{glmnet} package by \citet{elasticnet} for lasso regression. The ensemble includes bagged regression trees \citep{breiman1996bagging}, random forests \citep{ho1995random, breiman2001random}, lasso regression, support vector machines \citep{Boseretal1992, CortesVapnik1995}, and neural networks \citep{mcculloch1943logical, ripley2007pattern}. Because the ensemble searches for the best linear combination of multiple learners, it is expected to outperform any single method such as lasso. Nevertheless, lasso remains useful because it provides interpretable coefficient estimates that allow us to assess the influence of explanatory variables on the probability of submitting the lowest bid.

We randomly split the data into training (75\%) and test (25\%) samples. Model parameters are estimated using the training data, applying 10-fold cross-validation to select the optimal penalization parameter for lasso and the optimal weights for the ensemble. Out-of-sample performance is then evaluated on the test data. Model tuning and evaluation are based on accuracy, defined as the proportion of bids for which the observed rank matches the rank predicted by the algorithm. A bid is classified as first (lowest) if the predicted probability exceeds 0.5; otherwise, it is classified as second lowest. We repeat the random splitting, training, and testing procedure 20 times and compute the average accuracy across all bids, as well as separately for lowest and second-lowest bids, across repetitions. This procedure is implemented separately for the ensemble and for lasso regression.

Table \ref{ensemble1} reports the ensemble results, while Table \ref{lasso1} summarizes the lasso estimates. Overall, predictive accuracy is lower for lasso than for the ensemble, which is expected since the ensemble combines multiple algorithms to optimize performance. In Table \ref{ensemble1}, accuracy ranges from 69.6\% to 71.0\% for models including cost variables, with fairly balanced performance across lowest and second-lowest bids, though slightly higher for identifying winners. In contrast, Model 2, which relies only on contract allocation variables, performs substantially worse, with an accuracy of only 55.0\% and greater class imbalance. These findings highlight the dominant predictive role of cost variables in determining whether a bid ranks first or second.

Table \ref{lasso2} reports the absolute values of the estimated (non-standardized) lasso coefficients, which differ across models because each specification includes a different number of regressors. The absolute coefficients on \textit{LNCAP}, \textit{LNDIST}, and \textit{LNDIST\_SEC} are consistently larger than those for the contract allocation variables. This confirms the logit and ordered logit results, indicating that cost variables possess substantially higher predictive power than contract allocation variables.

{\renewcommand{\arraystretch}{1.1}
\begin{table} [!htp]
\begin{center}
\begin{tabular}{lrrr}
  \hline
Model & Accuracy & Acc. rank 1& Acc. rank 2  \\ 
  \hline
Model 1& 0.696 & 0.730 & 0.665 \\
Model 2& 0.550 & 0.675 & 0.428 \\ 
Model 3& 0.706 & 0.744 & 0.670 \\
Model 4& 0.710 & 0.753 & 0.669 \\
Model 5& 0.703 & 0.738 & 0.670 \\
Model 6& 0.707 & 0.742 & 0.675 \\
   \hline
\end{tabular}
\end{center}
\caption{Results ensemble method for training and testing with the bids of rank 1 and 2} \label{ensemble1}
\par
{\footnotesize Note:  Model, Accuracy, Acc. rank 1 and Acc. rank 2 denote the model estimated, the overall accuracy, the accuracy for classifying the bids of rank 1, and the accuracy for classifying the bids of rank 2, respectively.}
\end{table}}

{\renewcommand{\arraystretch}{1.1}
\begin{table} [!htp]
\begin{center}
\begin{tabular}{lrrr}
  \hline
Model & Accuracy & Acc. rank 1& Acc. rank 2  \\ 
  \hline
Model 1& 0.578 & 0.726 & 0.436 \\
Model 2& 0.497 & 0.927 & 0.067 \\ 
Model 3& 0.573 & 0.719 & 0.433 \\
Model 4& 0.555 & 0.736 & 0.375 \\
Model 5& 0.598 & 0.752 & 0.448 \\
Model 6& 0.589 & 0.754 & 0.424 \\
   \hline
\end{tabular}
\end{center}
\caption{Results lasso for training and testing with the bids of rank 1 and 2} \label{lasso1}
\par
{\footnotesize Note:  Model, Accuracy, Acc. rank 1 and Acc. rank 2 denote the model estimated, the overall accuracy, the accuracy for classifying the bids of rank 1, and the accuracy for classifying the bids of rank 2, respectively.}
\end{table}}

{\renewcommand{\arraystretch}{1.1}
\begin{table} [!htp]
\begin{center}
\begin{tabular}{lrrrrrr}
  \hline
Variable & Model 3& Model 4 & Model 5 & Model 6  \\ 
  \hline
LNCAP             & 0.2729 & 0.1994 & 0.2576 & 0.2345\\
LNDIST            & 0.3647 & 0.2853 & 0.3352 & 0.3083\\ 
LNCAP\_SEC  & 0.1159 & 0.0888 & 0.1145 & 0.0950\\
LNDIST\_SEC & 0.4313 & 0.3395 & 0.2684 & 0.2628\\ 
LASTWIN        & 0.0002 & 0.0001 & 0.0000 & 0.0000\\
SCORE           & 0.0248 & 0.0178 & 0.0291 & 0.0275\\
EQUAL            & 0.0030 & 0.0001 & 0.0006 & 0.0006\\
   \hline
\end{tabular}
\end{center}
\caption{Variable importance from lasso estimation} \label{lasso2}
\par
\end{table}}

\newpage

\setcounter{table}{0}
\renewcommand{\thetable}{C.\arabic{table}}
\section{Econometric tests}
\label{AppendixC}

For implementing the econometric tests, we estimate the following reduced-form bidding function:
\begin{equation}\label{eqestim2ch2}
LNBID_{it} = \beta_{0} + \gamma_{i} + \alpha_{t} 
+ \beta_{1,i} LNDIST_{it} 
+ \beta_{2,i} LNCAP_{it} 
+ \beta_{3,i} LNDIST\_SEC_{it} 
+ \beta_{4,i} LNCAP\_SEC_{it} 
+ \epsilon_{it},
\end{equation}
where $\alpha_{t}$ and $\gamma_{i}$ denote contract and firm fixed effects, respectively, and $\beta_{0}$ is the intercept. The explanatory variables proxy for cost determinants: \textit{LNDIST} is the logarithm of bidder distance to the project site; \textit{LNCAP} is the logarithm of the firm’s backlog; \textit{LNDIST\_SEC} is the logarithm of the minimum rival distance; and \textit{LNCAP\_SEC} is the logarithm of the minimum rival backlog. The slope coefficients $\beta_{1,i}$ to $\beta_{4,i}$ are allowed to vary across firms, which is essential for testing exchangeability. The term $\epsilon_{it}$ denotes the idiosyncratic error term.

To test conditional independence, we follow the procedure proposed by Bajari and Ye (2003). After estimating equation \ref{eqestim2ch2}, we recover the residuals $\hat{\epsilon}_{it}$ and examine whether they are correlated across firms participating in the same auction. The underlying idea is that, conditional on observed cost proxies, bids should be independent across firms in a competitive environment. Systematic correlation in residuals would instead indicate coordinated behavior beyond observable cost factors.

The procedure is implemented pairwise. First, we identify all firms in the sample and generate all possible firm pairs $(i,j)$. For each pair, we restrict attention to auctions in which both firms simultaneously submitted a bid and retain only these joint observations. The test is conducted only for firm pairs sharing at least five common auctions to ensure a sufficient number of observations.
For each admissible pair, we compute the Pearson correlation coefficient $\rho_{ij}$ between the residuals of firms $i$ and $j$. The corresponding test statistic is given by
\begin{equation}
t_{ij} = \rho_{ij} \sqrt{\frac{T_n - 2}{1 - \rho_{ij}^2}},
\end{equation}
where $T_n$ denotes the number of common auctions. Under the null hypothesis of conditional independence,
\begin{equation}
H_0: \rho_{ij} = 0,
\end{equation}
the statistic follows a Student-$t$ distribution with $T_n - 2$ degrees of freedom. We compute two-sided p-values for each firm pair.
For every tested pair, we report the firm identifiers, the estimated correlation coefficient, the test statistic, the degrees of freedom, the p-value, and the number of common auctions. 

The exchangeability test evaluates whether firms respond symmetrically to observable cost proxies. In a competitive environment, firms with identical cost realizations should submit comparable bids. Formally, exchangeability requires that the slope coefficients associated with the cost proxies be identical across firms. The null hypothesis can therefore be written as
\begin{equation}
H_{0}: \beta_{k i} = \beta_{k j} 
\quad \forall i \neq j, \quad \forall k = 1,\ldots,4,
\end{equation}
where $k$ indexes the four cost proxies included in equation \ref{eqestim2ch2}.

We implement this test pairwise for all firm combinations. For each pair $(i,j)$, we restrict attention to auctions in which both firms simultaneously submitted a bid and conduct the test only if the pair shares at least five common projects. Let $\boldsymbol{\beta}_i$ and $\boldsymbol{\beta}_j$ denote the vectors of firm-specific slope coefficients associated with the cost proxies. We test the joint null hypothesis $\boldsymbol{\beta}_i = \boldsymbol{\beta}_j$ using a Wald statistic of the form
\begin{equation}
W_{ij} = (\boldsymbol{\beta}_i - \boldsymbol{\beta}_j)' 
\mathbf{V}_{ij}^{-1}
(\boldsymbol{\beta}_i - \boldsymbol{\beta}_j),
\end{equation}
where $\mathbf{V}_{ij}$ is the heteroskedasticity-robust covariance matrix of the difference vector. The covariance matrix is constructed using White’s robust estimator applied to the full model.
Under the null hypothesis of exchangeability, $W_{ij}$ follows a $\chi^{2}$ distribution with degrees of freedom equal to the number of tested coefficients. We compute p-values accordingly. For each admissible firm pair, we report the Wald statistic, the degrees of freedom, the p-value, and the number of common projects.
For both tests, the procedure is implemented in \textsf{R}.

\subsection{Tests in the cartel period}
\begingroup
\setlength{\tabcolsep}{3pt}        
\renewcommand{\arraystretch}{0.85} 
\small                             

\begin{longtable}{rrrrrlrl}
\caption{Pairwise Bajari and Ye independence tests in the cartel period}
\label{tab:BY_pairwise_independence} \\

\toprule
firm\_i & firm\_j & corr & statistic & df & p.value & N.obs & p.adj \\
\midrule
\endfirsthead

\caption[]{Pairwise Bajari and Ye Independence Tests in the cartel period (continued)} \\
\toprule
firm\_i & firm\_j & corr & statistic & df & p.value & N.obs & p.adj \\
\midrule
\endhead

\midrule
\multicolumn{8}{r}{\textit{Continued on next page}} \\
\midrule
\endfoot

\bottomrule
\endlastfoot

5 &  15 & -0.59 & -4.41 &  36 & 0.0001 &  38 & 0.0092 \\ 
   11 &  12 & -0.82 & -4.68 &  11 & 0.0007 &  13 & 0.0346 \\ 
    6 &  12 & -0.76 & -4.09 &  12 & 0.0015 &  14 & 0.0513 \\ 
    4 &  14 & -0.63 & -3.55 &  19 & 0.0021 &  21 & 0.0548 \\ 
    6 &  13 & -0.32 & -2.75 &  68 & 0.0076 &  70 & 0.1564 \\ 
   13 &  15 & -0.36 & -2.44 &  40 & 0.0191 &  42 & 0.2556 \\ 
    7 &  12 & -0.68 & -2.82 &   9 & 0.0200 &  11 & 0.2556 \\ 
    1 &   8 & -0.57 & -2.53 &  13 & 0.0253 &  15 & 0.2556 \\ 
    2 &   6 & -0.25 & -2.27 &  77 & 0.0258 &  79 & 0.2556 \\ 
    2 &  13 & -0.29 & -2.25 &  56 & 0.0284 &  58 & 0.2556 \\ 
    5 &  11 & -0.29 & -2.24 &  54 & 0.0292 &  56 & 0.2556 \\ 
    8 &  14 & -0.49 & -2.35 &  17 & 0.0311 &  19 & 0.2556 \\ 
    1 &   4 & -0.67 & -2.54 &   8 & 0.0347 &  10 & 0.2556 \\ 
   10 &  11 & -0.34 & -2.19 &  38 & 0.0347 &  40 & 0.2556 \\ 
   12 &  15 & -0.68 & -2.43 &   7 & 0.0455 &   9 & 0.2936 \\ 
    1 &   6 & 0.48 & 2.14 &  15 & 0.0495 &  17 & 0.2936 \\ 
    2 &  12 & -0.58 & -2.23 &  10 & 0.0498 &  12 & 0.2936 \\ 
    5 &   6 & -0.25 & -1.97 &  57 & 0.0541 &  59 & 0.2936 \\ 
    5 &   7 & -0.34 & -2.00 &  30 & 0.0542 &  32 & 0.2936 \\ 
   10 &  13 & -0.34 & -1.96 &  30 & 0.0588 &  32 & 0.2936 \\ 
    8 &  10 & -0.41 & -1.99 &  20 & 0.0609 &  22 & 0.2936 \\ 
    5 &   8 & 0.61 & 2.16 &   8 & 0.0627 &  10 & 0.2936 \\ 
    2 &   8 & -0.39 & -1.92 &  21 & 0.0689 &  23 & 0.3020 \\ 
    9 &  11 & -0.43 & -1.92 &  16 & 0.0724 &  18 & 0.3020 \\ 
    2 &  14 & 0.34 & 1.87 &  26 & 0.0733 &  28 & 0.3020 \\ 
    8 &  11 & -0.41 & -1.87 &  17 & 0.0789 &  19 & 0.3119 \\ 
    3 &  13 & -0.65 & -2.09 &   6 & 0.0818 &   8 & 0.3119 \\ 
    6 &  10 & 0.24 & 1.71 &  46 & 0.0947 &  48 & 0.3482 \\ 
    1 &   5 & -0.51 & -1.79 &   9 & 0.1063 &  11 & 0.3741 \\ 
    4 &   8 & 0.38 & 1.69 &  17 & 0.1090 &  19 & 0.3741 \\ 
    7 &  16 & -0.36 & -1.60 &  17 & 0.1290 &  19 & 0.4285 \\ 
    8 &  16 & -0.53 & -1.64 &   7 & 0.1445 &   9 & 0.4652 \\ 
    7 &  13 & -0.24 & -1.44 &  33 & 0.1596 &  35 & 0.4981 \\ 
    4 &   6 & -0.26 & -1.42 &  27 & 0.1659 &  29 & 0.5026 \\ 
   10 &  15 & -0.29 & -1.41 &  21 & 0.1729 &  23 & 0.5087 \\ 
    2 &  11 & -0.16 & -1.32 &  65 & 0.1911 &  67 & 0.5408 \\ 
    2 &   7 & -0.20 & -1.31 &  41 & 0.1979 &  43 & 0.5408 \\ 
   11 &  14 & -0.28 & -1.32 &  21 & 0.2002 &  23 & 0.5408 \\ 
    7 &   8 & -0.44 & -1.38 &   8 & 0.2048 &  10 & 0.5408 \\ 
    5 &  13 & -0.18 & -1.27 &  50 & 0.2101 &  52 & 0.5410 \\ 
    7 &  10 & -0.28 & -1.25 &  19 & 0.2258 &  21 & 0.5485 \\ 
    1 &  14 & 0.26 & 1.22 &  20 & 0.2351 &  22 & 0.5485 \\ 
    7 &  11 & 0.17 & 1.20 &  46 & 0.2375 &  48 & 0.5485 \\ 
   11 &  13 & -0.15 & -1.19 &  64 & 0.2381 &  66 & 0.5485 \\ 
    1 &  16 & 0.56 & 1.34 &   4 & 0.2502 &   6 & 0.5485 \\ 
    3 &  11 & -0.38 & -1.23 &   9 & 0.2511 &  11 & 0.5485 \\ 
    4 &  10 & -0.23 & -1.16 &  24 & 0.2592 &  26 & 0.5485 \\ 
    6 &  11 & 0.12 & 1.13 &  91 & 0.2611 &  93 & 0.5485 \\ 
    4 &  15 & -0.26 & -1.16 &  18 & 0.2623 &  20 & 0.5485 \\ 
    2 &  15 & 0.18 & 1.13 &  40 & 0.2662 &  42 & 0.5485 \\ 
   13 &  14 & -0.24 & -1.12 &  20 & 0.2743 &  22 & 0.5540 \\ 
    1 &   7 & -0.32 & -1.08 &  10 & 0.3047 &  12 & 0.5926 \\ 
    3 &  15 & 0.49 & 1.14 &   4 & 0.3194 &   6 & 0.5926 \\ 
    4 &  12 & 0.37 & 1.07 &   7 & 0.3217 &   9 & 0.5926 \\ 
    4 &   5 & 0.27 & 1.02 &  13 & 0.3264 &  15 & 0.5926 \\ 
    4 &  13 & -0.20 & -0.99 &  25 & 0.3294 &  27 & 0.5926 \\ 
   12 &  16 & -0.48 & -1.11 &   4 & 0.3303 &   6 & 0.5926 \\ 
    9 &  13 & -0.28 & -1.01 &  12 & 0.3337 &  14 & 0.5926 \\ 
    5 &  14 & -0.28 & -0.94 &  10 & 0.3696 &  12 & 0.6452 \\ 
    3 &   6 & -0.26 & -0.89 &  11 & 0.3917 &  13 & 0.6501 \\ 
    5 &  10 & -0.18 & -0.87 &  22 & 0.3926 &  24 & 0.6501 \\ 
    4 &  16 & -0.25 & -0.88 &  12 & 0.3944 &  14 & 0.6501 \\ 
    6 &  16 & -0.15 & -0.85 &  32 & 0.4011 &  34 & 0.6501 \\ 
    2 &   4 & -0.14 & -0.85 &  34 & 0.4040 &  36 & 0.6501 \\ 
    8 &  15 & 0.28 & 0.81 &   8 & 0.4406 &  10 & 0.6962 \\ 
    4 &  11 & -0.16 & -0.77 &  24 & 0.4461 &  26 & 0.6962 \\ 
    3 &  16 & -0.38 & -0.83 &   4 & 0.4539 &   6 & 0.6964 \\ 
    1 &  13 & -0.20 & -0.76 &  14 & 0.4598 &  16 & 0.6964 \\ 
    6 &   8 & -0.17 & -0.73 &  19 & 0.4723 &  21 & 0.7020 \\ 
   12 &  13 & -0.24 & -0.74 &   9 & 0.4771 &  11 & 0.7020 \\ 
    9 &  15 & 0.36 & 0.77 &   4 & 0.4842 &   6 & 0.7024 \\ 
    6 &   7 & -0.10 & -0.68 &  48 & 0.4985 &  50 & 0.7132 \\ 
   10 &  14 & -0.12 & -0.63 &  28 & 0.5313 &  30 & 0.7496 \\ 
   10 &  12 & -0.25 & -0.62 &   6 & 0.5575 &   8 & 0.7688 \\ 
    5 &   9 & -0.24 & -0.62 &   6 & 0.5598 &   8 & 0.7688 \\ 
    6 &  14 & -0.11 & -0.56 &  24 & 0.5791 &  26 & 0.7807 \\ 
    4 &   7 & 0.15 & 0.55 &  13 & 0.5892 &  15 & 0.7807 \\ 
    6 &  15 & -0.07 & -0.53 &  52 & 0.5977 &  54 & 0.7807 \\ 
    1 &  15 & -0.26 & -0.55 &   4 & 0.6123 &   6 & 0.7807 \\ 
   11 &  15 & -0.07 & -0.51 &  47 & 0.6138 &  49 & 0.7807 \\ 
    1 &  11 & 0.14 & 0.51 &  14 & 0.6150 &  16 & 0.7807 \\ 
    2 &   5 & -0.07 & -0.50 &  46 & 0.6215 &  48 & 0.7807 \\ 
   14 &  15 & -0.13 & -0.40 &   9 & 0.6962 &  11 & 0.8615 \\ 
   11 &  16 & -0.07 & -0.39 &  29 & 0.7026 &  31 & 0.8615 \\ 
   10 &  16 & 0.10 & 0.37 &  14 & 0.7202 &  16 & 0.8728 \\ 
   12 &  14 & -0.13 & -0.28 &   5 & 0.7874 &   7 & 0.9431 \\ 
    8 &  12 & 0.13 & 0.25 &   4 & 0.8115 &   6 & 0.9566 \\ 
    2 &  16 & -0.04 & -0.22 &  31 & 0.8306 &  33 & 0.9566 \\ 
    2 &  10 & -0.03 & -0.21 &  40 & 0.8311 &  42 & 0.9566 \\ 
    5 &  12 & -0.07 & -0.21 &   8 & 0.8391 &  10 & 0.9566 \\ 
   14 &  16 & 0.06 & 0.18 &  10 & 0.8619 &  12 & 0.9566 \\ 
   15 &  16 & 0.04 & 0.18 &  19 & 0.8623 &  21 & 0.9566 \\ 
    8 &  13 & -0.04 & -0.17 &  16 & 0.8664 &  18 & 0.9566 \\ 
    5 &  16 & -0.03 & -0.15 &  22 & 0.8790 &  24 & 0.9566 \\ 
    3 &   9 & 0.05 & 0.15 &  11 & 0.8823 &  13 & 0.9566 \\ 
    9 &  16 & 0.07 & 0.12 &   3 & 0.9112 &   5 & 0.9761 \\ 
    7 &  15 & -0.02 & -0.10 &  35 & 0.9225 &  37 & 0.9761 \\ 
    6 &   9 & 0.02 & 0.09 &  19 & 0.9312 &  21 & 0.9761 \\ 
    1 &   2 & 0.02 & 0.07 &  16 & 0.9449 &  18 & 0.9761 \\ 
   13 &  16 & -0.01 & -0.07 &  29 & 0.9477 &  31 & 0.9761 \\ 
    7 &  14 & -0.01 & -0.04 &  13 & 0.9725 &  15 & 0.9824 \\ 
    1 &  10 & -0.01 & -0.03 &  18 & 0.9729 &  20 & 0.9824 \\ 
    1 &  12 & 0.01 & 0.01 &   3 & 0.9924 &   5 & 0.9924 \\ 
   \hline

\end{longtable}
\endgroup

\begingroup
\setlength{\tabcolsep}{3pt}       
\renewcommand{\arraystretch}{0.85} 
\small                             
\begin{longtable}{rrrrlrl}
\caption{Pairwise Exchangeability Tests on Bids in the cartel period}
\label{tab:exchangeability_pairwise} \\

\toprule
firm\_i & firm\_j & statistic & df & p.value & N.Obs & p.adj \\
\midrule
\endfirsthead

\caption[]{Pairwise Exchangeability Tests on Bids in the cartel period (continued)} \\
\toprule
firm\_i & firm\_j & statistic & df & p.value & N.Obs & p.adj \\
\midrule
\endhead

\midrule
\multicolumn{7}{r}{\textit{Continued on next page}} \\
\midrule
\endfoot

\bottomrule
\endlastfoot
   3 &  11 & 74.39 &   4 & 0.0000 &  11 & 0.0000 \\ 
    3 &   6 & 66.28 &   4 & 0.0000 &  13 & 0.0000 \\ 
    3 &  15 & 45.10 &   4 & 0.0000 &   6 & 0.0000 \\ 
    3 &  16 & 27.91 &   4 & 0.0000 &   6 & 0.0003 \\ 
    7 &  14 & 28.03 &   4 & 0.0000 &  15 & 0.0003 \\ 
    7 &  15 & 26.87 &   4 & 0.0000 &  37 & 0.0004 \\ 
    6 &   9 & 25.70 &   4 & 0.0000 &  21 & 0.0005 \\ 
    9 &  11 & 23.51 &   4 & 0.0001 &  18 & 0.0013 \\ 
    7 &  13 & 23.11 &   4 & 0.0001 &  35 & 0.0014 \\ 
   14 &  15 & 21.87 &   4 & 0.0002 &  11 & 0.0022 \\ 
    2 &  14 & 21.57 &   4 & 0.0002 &  28 & 0.0023 \\ 
    6 &  14 & 21.03 &   4 & 0.0003 &  26 & 0.0027 \\ 
    4 &  14 & 19.43 &   4 & 0.0006 &  21 & 0.0051 \\ 
   11 &  14 & 18.33 &   4 & 0.0011 &  23 & 0.0078 \\ 
    2 &  16 & 15.97 &   4 & 0.0031 &  33 & 0.0210 \\ 
    3 &  13 & 15.38 &   4 & 0.0040 &   8 & 0.0256 \\ 
   14 &  16 & 14.51 &   4 & 0.0058 &  12 & 0.0353 \\ 
    7 &  10 & 14.16 &   4 & 0.0068 &  21 & 0.0390 \\ 
    1 &   8 & 13.07 &   4 & 0.0109 &  15 & 0.0592 \\ 
   11 &  15 & 12.92 &   4 & 0.0117 &  49 & 0.0601 \\ 
    8 &  14 & 12.38 &   4 & 0.0147 &  19 & 0.0691 \\ 
    7 &  11 & 12.34 &   4 & 0.0150 &  48 & 0.0691 \\ 
    1 &   4 & 12.27 &   4 & 0.0154 &  10 & 0.0691 \\ 
   13 &  14 & 12.12 &   4 & 0.0165 &  22 & 0.0707 \\ 
    5 &  14 & 11.93 &   4 & 0.0179 &  12 & 0.0738 \\ 
    5 &   9 & 11.49 &   4 & 0.0216 &   8 & 0.0801 \\ 
    7 &   8 & 11.49 &   4 & 0.0216 &  10 & 0.0801 \\ 
    1 &   7 & 11.47 &   4 & 0.0218 &  12 & 0.0801 \\ 
    4 &  16 & 10.87 &   4 & 0.0281 &  14 & 0.0998 \\ 
    8 &  11 & 10.68 &   4 & 0.0304 &  19 & 0.1044 \\ 
    5 &   8 & 10.20 &   4 & 0.0372 &  10 & 0.1236 \\ 
    6 &   7 & 9.81 &   4 & 0.0438 &  50 & 0.1409 \\ 
    5 &  10 & 9.59 &   4 & 0.0479 &  24 & 0.1479 \\ 
   11 &  13 & 9.53 &   4 & 0.0491 &  66 & 0.1479 \\ 
    6 &  15 & 9.48 &   4 & 0.0502 &  54 & 0.1479 \\ 
    5 &   7 & 9.06 &   4 & 0.0596 &  32 & 0.1706 \\ 
    6 &  13 & 8.99 &   4 & 0.0613 &  70 & 0.1706 \\ 
    1 &   2 & 8.70 &   4 & 0.0690 &  18 & 0.1788 \\ 
   12 &  14 & 8.70 &   4 & 0.0691 &   7 & 0.1788 \\ 
    1 &   6 & 8.69 &   4 & 0.0694 &  17 & 0.1788 \\ 
    6 &   8 & 8.51 &   4 & 0.0747 &  21 & 0.1876 \\ 
    6 &  10 & 8.28 &   4 & 0.0819 &  48 & 0.1930 \\ 
    4 &   5 & 8.27 &   4 & 0.0821 &  15 & 0.1930 \\ 
    9 &  15 & 8.26 &   4 & 0.0824 &   6 & 0.1930 \\ 
    4 &   7 & 7.82 &   4 & 0.0982 &  15 & 0.2187 \\ 
   10 &  11 & 7.79 &   4 & 0.0997 &  40 & 0.2187 \\ 
    9 &  16 & 7.75 &   4 & 0.1012 &   5 & 0.2187 \\ 
    3 &   9 & 7.73 &   4 & 0.1019 &  13 & 0.2187 \\ 
    2 &   7 & 7.38 &   4 & 0.1172 &  43 & 0.2464 \\ 
    1 &  16 & 7.32 &   4 & 0.1197 &   6 & 0.2466 \\ 
    4 &  10 & 7.17 &   4 & 0.1273 &  26 & 0.2553 \\ 
    2 &  10 & 7.14 &   4 & 0.1289 &  42 & 0.2553 \\ 
    1 &  11 & 7.05 &   4 & 0.1333 &  16 & 0.2591 \\ 
   10 &  16 & 6.94 &   4 & 0.1389 &  16 & 0.2618 \\ 
    8 &  16 & 6.93 &   4 & 0.1398 &   9 & 0.2618 \\ 
    9 &  13 & 6.65 &   4 & 0.1557 &  14 & 0.2864 \\ 
    1 &  10 & 6.46 &   4 & 0.1675 &  20 & 0.3026 \\ 
   10 &  15 & 6.18 &   4 & 0.1860 &  23 & 0.3303 \\ 
    4 &   8 & 6.03 &   4 & 0.1971 &  19 & 0.3440 \\ 
    4 &  15 & 5.92 &   4 & 0.2053 &  20 & 0.3511 \\ 
    2 &   8 & 5.88 &   4 & 0.2079 &  23 & 0.3511 \\ 
    2 &  15 & 5.75 &   4 & 0.2187 &  42 & 0.3633 \\ 
    1 &  15 & 5.70 &   4 & 0.2228 &   6 & 0.3643 \\ 
    6 &  16 & 5.61 &   4 & 0.2300 &  34 & 0.3671 \\ 
    5 &  16 & 5.57 &   4 & 0.2338 &  24 & 0.3671 \\ 
   11 &  16 & 5.55 &   4 & 0.2352 &  31 & 0.3671 \\ 
    5 &   6 & 5.27 &   4 & 0.2606 &  59 & 0.3974 \\ 
    8 &  10 & 5.25 &   4 & 0.2624 &  22 & 0.3974 \\ 
    4 &  13 & 4.88 &   4 & 0.2996 &  27 & 0.4472 \\ 
    5 &  11 & 4.57 &   4 & 0.3345 &  56 & 0.4922 \\ 
    7 &  16 & 4.37 &   4 & 0.3577 &  19 & 0.5189 \\ 
    1 &  14 & 4.22 &   4 & 0.3771 &  22 & 0.5321 \\ 
   10 &  14 & 4.22 &   4 & 0.3771 &  30 & 0.5321 \\ 
    1 &   5 & 4.14 &   4 & 0.3871 &  11 & 0.5388 \\ 
    2 &  13 & 4.03 &   4 & 0.4015 &  58 & 0.5513 \\ 
    4 &  11 & 3.81 &   4 & 0.4329 &  26 & 0.5815 \\ 
    2 &  11 & 3.79 &   4 & 0.4348 &  67 & 0.5815 \\ 
    1 &  12 & 3.69 &   4 & 0.4499 &   5 & 0.5867 \\ 
    2 &   5 & 3.69 &   4 & 0.4500 &  48 & 0.5867 \\ 
    7 &  12 & 3.47 &   4 & 0.4829 &  11 & 0.6218 \\ 
   10 &  12 & 3.37 &   4 & 0.4974 &   8 & 0.6325 \\ 
    1 &  13 & 3.33 &   4 & 0.5038 &  16 & 0.6328 \\ 
   10 &  13 & 3.19 &   4 & 0.5265 &  32 & 0.6479 \\ 
    2 &   6 & 3.18 &   4 & 0.5284 &  79 & 0.6479 \\ 
    4 &   6 & 3.13 &   4 & 0.5355 &  29 & 0.6489 \\ 
    5 &  13 & 2.37 &   4 & 0.6686 &  52 & 0.8008 \\ 
    8 &  12 & 2.14 &   4 & 0.7109 &   6 & 0.8413 \\ 
    8 &  15 & 2.09 &   4 & 0.7187 &  10 & 0.8413 \\ 
    4 &  12 & 1.90 &   4 & 0.7549 &   9 & 0.8737 \\ 
    5 &  15 & 1.72 &   4 & 0.7876 &  38 & 0.8840 \\ 
   15 &  16 & 1.69 &   4 & 0.7925 &  21 & 0.8840 \\ 
    2 &   4 & 1.69 &   4 & 0.7932 &  36 & 0.8840 \\ 
   12 &  16 & 1.61 &   4 & 0.8062 &   6 & 0.8840 \\ 
   13 &  15 & 1.61 &   4 & 0.8068 &  42 & 0.8840 \\ 
   12 &  15 & 1.34 &   4 & 0.8539 &   9 & 0.9195 \\ 
   13 &  16 & 1.31 &   4 & 0.8597 &  31 & 0.9195 \\ 
    2 &  12 & 1.26 &   4 & 0.8674 &  12 & 0.9195 \\ 
   12 &  13 & 1.19 &   4 & 0.8798 &  11 & 0.9195 \\ 
   11 &  12 & 1.09 &   4 & 0.8965 &  13 & 0.9195 \\ 
    6 &  12 & 1.06 &   4 & 0.9001 &  14 & 0.9195 \\ 
    5 &  12 & 1.02 &   4 & 0.9065 &  10 & 0.9195 \\ 
    8 &  13 & 1.00 &   4 & 0.9105 &  18 & 0.9195 \\ 
    6 &  11 & 0.33 &   4 & 0.9881 &  93 & 0.9881 \\ 
   \hline
\end{longtable}
\endgroup

\subsection{Tests in the post-cartel period}
\begingroup
\setlength{\tabcolsep}{3pt}        
\renewcommand{\arraystretch}{0.85} 
\small                            

\begin{longtable}{rrrrrlrl}
\caption{Pairwise Bajari and Ye independence tests in the post-cartel period}
\label{tab:BY_pairwise_independence} \\

\toprule
firm\_i & firm\_j & corr & statistic & df & p.value & N.obs & p.adj \\
\midrule
\endfirsthead

\caption[]{Pairwise Bajari and Ye Independence Tests in the post-cartel period (continued)} \\
\toprule
firm\_i & firm\_j & corr & statistic & df & p.value & N.obs & p.adj \\
\midrule
\endhead

\midrule
\multicolumn{8}{r}{\textit{Continued on next page}} \\
\midrule
\endfoot

\bottomrule
\endlastfoot

2 &   4 & -0.90 & -5.39 &   7 & 0.0010 &   9 & 0.0480 \\ 
    6 &  11 & -0.62 & -3.16 &  16 & 0.0061 &  18 & 0.1104 \\ 
   11 &  15 & -0.73 & -3.38 &  10 & 0.0070 &  12 & 0.1104 \\ 
    2 &   6 & -0.61 & -2.58 &  11 & 0.0258 &  13 & 0.3028 \\ 
    6 &  12 & 0.78 & 2.52 &   4 & 0.0652 &   6 & 0.6129 \\ 
    6 &   7 & -0.49 & -1.78 &  10 & 0.1054 &  12 & 0.6480 \\ 
    6 &  14 & 0.78 & 2.13 &   3 & 0.1231 &   5 & 0.6480 \\ 
   12 &  15 & -0.68 & -1.85 &   4 & 0.1383 &   6 & 0.6480 \\ 
    2 &   5 & -0.53 & -1.65 &   7 & 0.1426 &   9 & 0.6480 \\ 
    6 &  10 & -0.53 & -1.65 &   7 & 0.1437 &   9 & 0.6480 \\ 
    7 &  10 & -0.74 & -1.91 &   3 & 0.1517 &   5 & 0.6480 \\ 
    5 &  13 & -0.45 & -1.50 &   9 & 0.1683 &  11 & 0.6591 \\ 
   12 &  13 & -0.52 & -1.47 &   6 & 0.1910 &   8 & 0.6905 \\ 
    7 &  15 & -0.48 & -1.33 &   6 & 0.2329 &   8 & 0.7820 \\ 
   10 &  11 & 0.58 & 1.23 &   3 & 0.3055 &   5 & 0.8697 \\ 
    4 &  11 & -0.44 & -1.10 &   5 & 0.3196 &   7 & 0.8697 \\ 
    4 &  15 & 0.40 & 1.06 &   6 & 0.3284 &   8 & 0.8697 \\ 
    4 &   5 & 0.47 & 1.07 &   4 & 0.3461 &   6 & 0.8697 \\ 
    5 &  15 & -0.34 & -0.89 &   6 & 0.4068 &   8 & 0.8697 \\ 
   11 &  12 & 0.48 & 0.96 &   3 & 0.4090 &   5 & 0.8697 \\ 
    2 &  10 & -0.41 & -0.90 &   4 & 0.4188 &   6 & 0.8697 \\ 
   13 &  14 & -0.45 & -0.87 &   3 & 0.4474 &   5 & 0.8697 \\ 
    5 &   6 & 0.26 & 0.79 &   9 & 0.4491 &  11 & 0.8697 \\ 
   10 &  13 & -0.27 & -0.74 &   7 & 0.4858 &   9 & 0.8697 \\ 
    4 &   7 & 0.26 & 0.72 &   7 & 0.4940 &   9 & 0.8697 \\ 
    4 &  10 & -0.30 & -0.70 &   5 & 0.5133 &   7 & 0.8697 \\ 
    6 &  13 & -0.15 & -0.60 &  16 & 0.5592 &  18 & 0.8697 \\ 
    4 &  12 & 0.29 & 0.61 &   4 & 0.5728 &   6 & 0.8697 \\ 
   10 &  14 & -0.34 & -0.63 &   3 & 0.5760 &   5 & 0.8697 \\ 
   10 &  15 & 0.30 & 0.55 &   3 & 0.6223 &   5 & 0.8697 \\ 
    5 &   7 & -0.23 & -0.52 &   5 & 0.6268 &   7 & 0.8697 \\ 
    5 &  11 & -0.17 & -0.47 &   7 & 0.6519 &   9 & 0.8697 \\ 
    5 &  12 & 0.27 & 0.49 &   3 & 0.6562 &   5 & 0.8697 \\ 
    4 &   6 & -0.14 & -0.45 &  10 & 0.6590 &  12 & 0.8697 \\ 
    2 &  12 & -0.27 & -0.48 &   3 & 0.6610 &   5 & 0.8697 \\ 
    2 &  13 & 0.13 & 0.43 &  11 & 0.6764 &  13 & 0.8697 \\ 
   10 &  12 & -0.19 & -0.43 &   5 & 0.6846 &   7 & 0.8697 \\ 
    4 &  13 & -0.12 & -0.38 &   9 & 0.7144 &  11 & 0.8836 \\ 
    2 &  15 & -0.14 & -0.34 &   6 & 0.7425 &   8 & 0.8872 \\ 
    2 &  11 & 0.11 & 0.32 &   8 & 0.7566 &  10 & 0.8872 \\ 
    2 &  14 & 0.18 & 0.31 &   3 & 0.7739 &   5 & 0.8872 \\ 
   11 &  13 & -0.07 & -0.25 &  13 & 0.8098 &  15 & 0.8882 \\ 
   13 &  15 & -0.07 & -0.22 &  11 & 0.8291 &  13 & 0.8882 \\ 
    7 &  11 & -0.11 & -0.23 &   4 & 0.8315 &   6 & 0.8882 \\ 
    7 &  13 & -0.04 & -0.12 &   8 & 0.9042 &  10 & 0.9444 \\ 
    6 &  15 & -0.02 & -0.08 &  10 & 0.9386 &  12 & 0.9590 \\ 
    2 &   7 & -0.01 & -0.03 &   6 & 0.9794 &   8 & 0.9794 \\ 
    \hline

\end{longtable}
\endgroup

\begingroup
\setlength{\tabcolsep}{3pt}        
\renewcommand{\arraystretch}{0.85} 
\small                             
\begin{longtable}{rrrrlrl}
\caption{Pairwise Exchangeability Tests on Bids in the post-cartel period}
\label{tab:exchangeability_pairwise} \\

\toprule
firm\_i & firm\_j & statistic & df & p.value & N.Obs & p.adj \\
\midrule
\endfirsthead

\caption[]{Pairwise Exchangeability Tests on Bids in the post-cartel period (continued)} \\
\toprule
firm\_i & firm\_j & statistic & df & p.value & N.Obs & p.adj \\
\midrule
\endhead

\midrule
\multicolumn{7}{r}{\textit{Continued on next page}} \\
\midrule
\endfoot

\bottomrule
\endlastfoot
 2 &  10 & 34.51 &   4 & 0.0000 &   6 & 0.0000 \\ 
    4 &  12 & 33.60 &   4 & 0.0000 &   6 & 0.0000 \\ 
   10 &  12 & 30.29 &   4 & 0.0000 &   7 & 0.0000 \\ 
    2 &  12 & 28.70 &   4 & 0.0000 &   5 & 0.0001 \\ 
    5 &  12 & 26.15 &   4 & 0.0000 &   5 & 0.0002 \\ 
   11 &  12 & 22.17 &   4 & 0.0002 &   5 & 0.0008 \\ 
    6 &  12 & 19.72 &   4 & 0.0006 &   6 & 0.0021 \\ 
    2 &   4 & 18.00 &   4 & 0.0012 &   9 & 0.0036 \\ 
    4 &   7 & 18.08 &   4 & 0.0012 &   9 & 0.0036 \\ 
    2 &   5 & 14.06 &   4 & 0.0071 &   9 & 0.0185 \\ 
    4 &   5 & 11.11 &   4 & 0.0254 &   6 & 0.0600 \\ 
    4 &  11 & 10.78 &   4 & 0.0292 &   7 & 0.0632 \\ 
    2 &   7 & 10.48 &   4 & 0.0331 &   8 & 0.0633 \\ 
    2 &   6 & 10.41 &   4 & 0.0341 &  13 & 0.0633 \\ 
    6 &   7 & 9.48 &   4 & 0.0501 &  12 & 0.0868 \\ 
    4 &  10 & 8.11 &   4 & 0.0875 &   7 & 0.1422 \\ 
    4 &   6 & 7.94 &   4 & 0.0937 &  12 & 0.1433 \\ 
    5 &   7 & 7.35 &   4 & 0.1185 &   7 & 0.1711 \\ 
    5 &  11 & 6.87 &   4 & 0.1430 &   9 & 0.1957 \\ 
    5 &   6 & 6.05 &   4 & 0.1958 &  11 & 0.2545 \\ 
   10 &  11 & 5.25 &   4 & 0.2623 &   5 & 0.3248 \\ 
    7 &  11 & 4.70 &   4 & 0.3195 &   6 & 0.3776 \\ 
    2 &  11 & 4.51 &   4 & 0.3410 &  10 & 0.3854 \\ 
    7 &  10 & 3.68 &   4 & 0.4506 &   5 & 0.4882 \\ 
    6 &  11 & 2.81 &   4 & 0.5900 &  18 & 0.6136 \\ 
    6 &  10 & 1.16 &   4 & 0.8854 &   9 & 0.8854 \\  
    \hline
\end{longtable}
\endgroup

\begin{spacing}{1.0}
\bibliographystyle{agu}
\bibliography {bibliothesis}
\end{spacing}

\end{document}